\documentclass[twocolumn,twocolappendix]{aastex631}

\usepackage{amsmath,amsfonts,amssymb,mathtools}
\usepackage{amsthm}
\usepackage{graphicx}

\usepackage{appendix}
\usepackage{tabularx}
\usepackage{float}
\usepackage{hyperref}
\usepackage{xcolor}

\usepackage{amsmath}
\usepackage{hyperref}
\usepackage{enumitem}
\usepackage{footnote}
\usepackage{tablefootnote}

\usepackage{lipsum}
\usepackage{comment}
\usepackage{multirow}
\usepackage{enumitem}
\usepackage{hyperref}
\usepackage{changepage}

\usepackage{xcolor, soul}

\newcommand{\afe}{$\rm [\alpha/Fe]$}
\newcommand{\feh}{$\rm [Fe/H]$}
\newcommand{\zh}{$\rm [Z/H]$}
\newcommand{\mgfe}{$\rm [Mg/Fe]$}

\newcommand{\newmist}{\texttt{MIST v2.3}}
\newcommand{\oldmist}{\texttt{MIST v1.2}}
\newcommand{\newctk}{\texttt{C3K v2.3}}
\newcommand{\ctk}{\texttt{C3K}}
\newcommand{\basti}{\texttt{BaSTI}}
\newcommand{\mist}{\texttt{MIST}}
\newcommand{\alf}{\texttt{alf}}
\newcommand{\smiles}{\texttt{sMILES}}
\newcommand{\amc}{\texttt{$\alpha$-MC}}
\newcommand{\mgb}{Mg \textit{b}}

\usepackage{natbib}
\defcitealias{GS98}{GS98}
\defcitealias{Coelho2007}{C07}

\graphicspath{{./}}

\begin{document}

\title{$\alpha$-MC: Self-consistent $\alpha$-enhanced stellar population models covering\\a wide range of age, metallicity, and wavelength}

\author[0000-0002-8435-9402]{Minjung Park}
\affiliation{Center for Astrophysics $\mid$ Harvard $\&$ Smithsonian, Cambridge, MA, USA}

\author{Charlie Conroy}
\affiliation{Center for Astrophysics $\mid$ Harvard $\&$ Smithsonian, Cambridge, MA, USA}

\author{Benjamin D. Johnson}
\affiliation{Center for Astrophysics $\mid$ Harvard $\&$ Smithsonian, Cambridge, MA, USA}

\author{Joel Leja}
\affiliation{Department of Astronomy \& Astrophysics, The Pennsylvania State University, University Park, PA, USA}
\affiliation{Institute for Gravitation and the Cosmos, The Pennsylvania State University, University Park, PA, USA}
\affiliation{Institute for Computational \& Data Sciences, The Pennsylvania State University, University Park, PA, USA}

\author{Aaron Dotter}
\affiliation{Department of Physics and Astronomy, Dartmouth College, 6127 Wilder Laboratory, Hanover, NH 03755, USA}

\author{Phillip A. Cargile}
\affiliation{Center for Astrophysics $\mid$ Harvard $\&$ Smithsonian, Cambridge, MA, USA}

\begin{abstract}

We present new stellar population models, $\alpha$-MC, self-consistently taking into account non-solar $\rm [\alpha/Fe]$ abundances for both isochrones and stellar spectra. 
The $\alpha$-MC models are based on $\alpha$-enhanced MIST isochrones and C3K spectral libraries, which are publicly available in FSPS. 
Our new models cover a wide range of ages ($\rm \log (age/yr) = 5.0 - 10.3$), metallicities ($\rm [Fe/H]=[-2.5,+0.5]$ in steps of 0.25, $\rm [\alpha/Fe]=-0.2,+0.0,+0.2,+0.4,+0.6$), and wavelengths ($0.1-2.5\,\rm \mu m$). 
We investigate the separate and combined effects of $\alpha$-enhanced isochrones and stellar spectral libraries on simple stellar populations (SSPs), including their broadband colors, spectral indices, and full spectra.
We find that the primary effect of $\alpha$-enhancement in isochrones is to lower the overall continuum levels and redden the continuum shapes, while $\alpha$-enhancement in stellar spectra mainly affects individual spectral lines. 
At constant $\rm [Fe/H]$, $\alpha$-enhancement has significant impacts on the broadband colors by $\rm \sim 0.1-0.4\,mag$ across all ages ($\rm 0.01 - 10\,Gyr$). 
The effects of $\alpha$-enhancement on colors at fixed $\rm [Z/H]$ are smaller, by $\rm \sim 0.1-0.2\,mag$.
The spectral indices involving $\alpha$-elements, Ca4227 and Mg \textit{b}, increase with $\rm [\alpha/Fe]$ (both at fixed $\rm [Fe/H]$ and fixed $\rm [Z/H]$) due to enhanced $\alpha$-abundances. 
At constant $\rm [Fe/H]$, $\alpha$-enhancement weakens most Fe-sensitive and Hydrogen Balmer lines. 
Our new self-consistent $\alpha$-enhanced models will be essential in deriving accurate physical properties of high-redshift galaxies, where $\alpha$-enhancement is expected to be common.

\end{abstract}


\section{Introduction} \label{sec:intro}
Modeling the spectral energy distribution (SED) of unresolved stellar systems, known as stellar population synthesis, is a powerful tool to infer their physical properties, such as ages, masses, and metallicity.  
The three main ingredients of a simple stellar population (with a single age and metallicity; SSP) are initial mass function (IMF), stellar isochrones, and stellar spectral libraries, each with uncertainties \citep[see the review of][and the references therein]{Walcher2011, Conroy2013}. 
Although numerous efforts have been made to improve the model ingredients, one of the unresolved limitations of currently available stellar population models is that most assume solar-scaled chemical abundance patterns. 
Specifically, most assume solar-scaled abundance ratio between $\alpha$ elements (e.g., O, Mg, Si, and Ca) and Fe-peak elements (i.e., \afe$=+0.0$). 
Such models have limited applicability to galaxies with non-solar abundance patterns (i.e., non-solar \afe).

The formation history of galaxies is imprinted on their stellar \afe, as $\alpha$ elements and Fe-peak elements are enriched by different sources on different timescales. 
The $\alpha$ elements are mostly enriched by core-collapse SNe (SNe II) on a short timescale ($\sim20\,\rm Myr$), whereas Fe-peak elements are enriched both by SNe II and SNe Ia with a delayed timescale ($\sim1\,\rm Gyr$). 
Thus, the abundance ratio of \afe\ is set by the relative importance of SNe II and SNe Ia, and \afe\ has been used as an independent probe for the formation timescales of galaxies \citep[e.g.,][]{Matteucci1994, Trager2000, Thomas2003, Thomas2005}.
At high redshifts, there was less time for SNe Ia, and SNe II were more dominant for chemical enrichment. 
Galaxies at high redshifts are, therefore, expected to be $\alpha$-enhanced (\afe$>0.0$), and it is important to take $\alpha$-enhancement into account to model their SEDs more accurately. 
Indeed, deep spectroscopy of massive quiescent galaxies at $z>2$ has revealed enhanced \mgfe, supporting their short formation timescales \citep[e.g.,][]{Kriek2016Nature, Kriek2019, Jafariyazani2020ApJ, Jafariyazani2024, Beverage2024_heavymetal, Beverage2024_suspense, Carnall2024excels}. 
Some studies have derived [O/Fe] of star-forming galaxies at $z\sim2$ by combining the gas-phase oxygen abundance and stellar iron abundance. 
For example, \cite{Steidel2016} have measured oxygen abundance $Z_{\rm neb}/Z_\odot\approx0.5$ from nebular emission and stellar metallicity $Z_{\rm star}/Z_\odot\approx0.1$ from FUV continuum (tracing stellar Fe abundances) and derived stellar O/Fe $\rm \approx4-5\,(O/Fe)_{\odot}$ for star-forming galaxies at $z\sim 2.4$ \citep[see also][]{Cullen2019, Topping2020, Kashino2022, Chartab2024_Latis}. 
However, measuring both stellar oxygen (or $\alpha$ abundances) and iron abundances in a consistent way requires a new stellar population model with $\alpha$-enhancement, incorporating both $\alpha$-enhanced stellar evolution (isochrones) and $\alpha$-enhanced stellar spectra.

Several studies have explored the effect of $\alpha$-enhancement on overall stellar evolution and have constructed isochrones with non-solar \afe\ abundances \citep[e.g.,][]{Salaris1993, Kim2002_Y2, Pietrinferni2006, Pietrinferni2021, Dotter2008, VandenBerg2014}. 
$\alpha$-enhancement has various impacts on stellar evolution, including the effective temperatures of the main-sequence turnoff and red giant stars, their lifetimes, opacities, and reaction rates of the CNO cycle. 
Some studies have further investigated the effect of individual elemental variations on stellar evolution \citep[e.g.,][]{Dotter2007}. 
Many of these studies have constructed $\alpha$-enhanced isochrones with the goal of understanding old stellar populations, such as globular clusters \citep[e.g.,][]{Kim2002_Y2, Lee2009_GCs}, galactic halo and bulge stars. 
Thus, they have focused mainly on the impacts on lower-mass stellar evolution (typically, up to $\sim10\,M_\odot$) and constructed $\alpha$-enhanced isochrones only for older ages (typically, $>1\,\rm Gyr$). 
To understand young stellar populations of galaxies in the early Universe, we need stellar evolution models assuming non-solar $\alpha$ abundances for higher-mass stars (and the resulting $\alpha$-enhanced isochrones at younger ages).

$\alpha$-enhancement also has significant effects on stellar spectra, which has been investigated in several previous studies \citep[e.g.,][]{Barbuy1994, Tripicco1995, Barbuy2003, Thomas2003, Korn2005, Serven2005}. 
There are two approaches to producing stellar spectral libraries with non-solar abundances. 
One approach is to use purely theoretical synthetic spectra.
For example, \cite{Coelho2005} (updated in \citealt{Coelho2014}) and \cite{AllendePrieto2018} have generated synthetic spectra with non-solar \afe abundances by performing radiative transfer through model stellar atmospheres. 
Synthetic spectral models have the advantage of wide stellar parameter and wavelength coverage. 
However, theoretical models have many well-known uncertainties including the 1D and LTE assumptions in model atmosphere calculations, incomplete line lists, and adopted microturbulence velocities. 
These uncertainties highlight the need for further testing to ensure that these synthetic models can reproduce observations.

An alternative approach is to apply only differential predictions from theoretical models to observed empirical spectra with solar abundance patterns \citep[e.g., ][]{Trager2000, Thomas2005, Serven2005, Walcher2009, CvD12}.  
For example, \cite{Conroy2018_alf} (\alf) have generated a ``response function’’ of each element (i.e., the fractional change in spectrum based on the change in individual chemical abundance) and applied this to the empirical spectral libraries to generate models that can measure individual element abundances. 
However, it is not entirely clear whether the effect of enhancing all $\alpha$ elements simultaneously is equivalent to the linear combination of the effect of individual elements. 
This is because the effect of the overall enhancement of all $\alpha$ elements on the continuum is different from the combined effect of individual elements on the continuum. 
Recently, \cite{Knowles2021} (\smiles) have produced semi-empirical stellar spectra with non-solar \afe\ ratios by applying the differential predictions to individual MILES stars, where they used \mgfe\ measurements of MILES stars \citep{Milone2011} as a proxy of \afe. 
However, semi-empirical models have limitations in that they are still dependent on the stellar parameters and wavelength ranges of the empirical spectra.

There have been some self-consistent models employing both $\alpha$-enhanced isochrones and stellar spectra. 
Earlier studies of \cite{Coelho2007}, \cite{Lee2009}, and \cite{Percival2009} have generated $\alpha$-enhanced SSPs by combining $\alpha$-enhanced isochrones with purely theoretical $\alpha$-enhanced spectra. 
\cite{Vazdekis2015} have used BASTI $\alpha$-enhanced isochrones \citep{Pietrinferni2006} and empirical MILES stellar spectra \citep{Sanchez-Blazquez2006}.
They then corrected their SSPs to synthesize $\alpha$-enhanced SSPs using a differential correction and \mgfe\ estimates of MILES stars from \citep{Milone2011} as a proxy of \afe. 
A recent study of \cite{Knowles2023} has also used the \mgfe\ estimates of MILES stars to correct the solar-scaled empirical spectra using a differential correction. 
The key difference is that they used a semi-empirical sMILES spectral library \citep{Knowles2021} where individual MILES stellar spectra are corrected before synthesizing. 
Although significant progress has been made in building self-consistent $\alpha$-enhanced SSP models, there are still several limitations in these models. 
In the case of semi-empirical models, they still suffer from limited stellar parameters and wavelength coverage. 
Also, most of these $\alpha$-enhanced isochrone models are computed only for older ages, thus, SSPs constructed from these models can only be applied to old stellar populations (typically $>1\,\rm Gyr$).

In this work, we present new self-consistent stellar population models with non-solar \afe\ ratios and compare them with other self-consistent $\alpha$-enhanced models. 
Applications to observed data will be presented in a subsequent paper. 
The new models are constructed from the \newmist\ isochrones (Dotter et al. in prep) and the new \ctk\ theoretical spectral library. 
For varying $\alpha$-abundances, all $\alpha$ elements (O, Ne, Mg, Si, S, Ar, Ca, and Ti) are scaled up/down uniformly and simultaneously. 
Our new \amc\ SSP models cover a wide range of ages ($\rm \log (age/yr) = 5.0 - 10.3$), metallicity (\feh$=[-2.5, +0.5]$ in steps of 0.25, and \afe$=-0.2,+0.0,+0.2,+0.4,+0.6$), and wavelength ($0.1-2.5\,\rm \mu m$). 
The $\alpha$-enhanced isochrones and stellar spectra share the same abundance mixtures from \cite{GS98}, thus our SSPs are fully consistent in terms of the abundances.

This paper is organized as follows.
In Section~\ref{sec:model_ingredients}, we describe the model ingredients of our new $\alpha$-enhanced SSP models.
In Section~\ref{sec:result_compare_old_mist_new_mist}, we compare SSPs generated with the new \newmist\ and \oldmist\ isochrones at solar abundances (\afe$=+0.0$). 
In Section~\ref{sec:result_afsps_compare_solar_vs_afe}, we investigate the effects of $\alpha$-enhancement on SSP spectra, broadband colors, and spectral indices. We also separate $\alpha$-enhancement effects of stellar evolution (isochrones) vs. stellar spectra. 
In Section~\ref{sec:result_compare_our_vs_other_afe_models}, we compare our \amc\ models with those from other self-consistent $\alpha$-enhanced models. 
We present the summary and our conclusion in Section~\ref{sec:summary_conclusion}.

\section{Model ingredients} \label{sec:model_ingredients}
To build a model of the integrated light from SSP, three main ingredients are required: the initial mass function (IMF), stellar isochrones, and stellar spectra. 
In this section we describe the new $\alpha$-enhanced isochrones and stellar spectra, and how we build $\alpha$-enhanced SSP models.

\subsection{\newmist\ isochrones} \label{sec:model_mistv2.3}
In this work, we use the \newmist\ isochrones, an upgraded version of \oldmist\ presented in \cite{Choi2016, Dotter2016}. 
The \mist\ isochrones are constructed based on the 1D stellar evolution code \texttt{MESA} \citep{Paxton2011} which computes the stellar evolutionary sequences for stars in the mass range of $0.1-300\,M_\odot$. 
The evolutionary sequences are tracked from pre-MS through the end of carbon burning (for high-mass stars, $M \geq 8\,M_\odot$) or the end of the AGB phases (for lower-mass stars $M < 8\,M_\odot$). 
The detailed physics adopted in the \mist\ models can be found in \cite{Choi2016}.

One of the major improvements of \newmist\ over \texttt{v1.2} is the inclusion of non-solar $\alpha$-abundances (variable \afe). 
The varying $\alpha$ abundances (\afe$=-0.2,+0.0,+0.2,+0.4,+0.6$) are computed in lockstep by scaling up/down the abundance of all $\alpha$ elements (O, Ne, Mg, Si, S, Ar, Ca, and Ti) uniformly at fixed \feh. 
The enhancement in all $\alpha$ elements affects the opacity, nuclear reaction rates, and equation of states throughout the evolutionary phases. 
We use two different radiative opacity tables depending on temperature: the \cite{Ferguson2005} opacity table for low temperatures ($T<10^4\,\rm K$) and the OPAL opacity tables \citep{Iglesias1993,Iglesias1996} for high temperatures ($T>10^4\,\rm K$).
The opacities for some of the most metal-rich grids (\feh, \afe) are not provided by these tables; thus, for these metal-rich grids, the isochrones for $M=0.1-0.5\,\rm M_\odot$ are copied from the isochrones of the nearest metallicity as follows:
\begin{itemize}[leftmargin=*]
\setlength\itemsep{-0.2em} 
    \item ($+0.25$,$+0.0$) is copied to ($+0.50$,$+0.0$)
    \item ($+0.25$,$+0.2$) is copied to ($+0.50$,$+0.2$)
    \item ($+0.00$,$+0.4$) is copied to ($+0.25$,$+0.4$), ($+0.50$,$+0.4$)
    \item ($-0.25$,$+0.6$) is copied to ($+0.00$,$+0.6$), ($+0.25$,$+0.6$)
\end{itemize}
For the most metal-rich grid ($+0.50$,$+0.6$), the entire isochrone data (all masses) are copied from the nearest grid of ($+0.50$,$+0.4$).

We adopt the solar abundances from \cite{GS98} (GS98; hereafter), following the discussions in \cite{Bergemann2014}. Traditionally, the solar abundance from \citetalias{GS98} has been widely used, which is derived from the analysis of the solar spectrum based on 1D model atmosphere calculations assuming LTE. 
However, non-LTE radiative transfer with 3D atmosphere models revealed a much lower solar abundance, which led to a major revision in the solar mixture as presented in \cite{Asplund2009}. 
Most recently, \cite{Magg2022} have analyzed the solar photospheric abundances and obtained $(Z/X) = 0.0225$, showing that the revised solar abundances are more consistent with the old values from \citetalias{GS98}, $(Z/X) = 0.023$. 
Therefore, we adopt the solar abundance from \citetalias{GS98}, as this abundance is very close to the recently revised value from \cite{Magg2022} and has been widely used in the literature.

\vspace{1cm}
Here we have given only a brief description of the new isochrone models. 
Full details about the new \newmist\ models will be provided in Dotter et al. in prep.

\subsection{\newctk\ spectral libraries}  \label{sec:model_c3kv2.3}

\begin{figure}[t!]
    \centering
    \includegraphics[width=1.0\columnwidth]{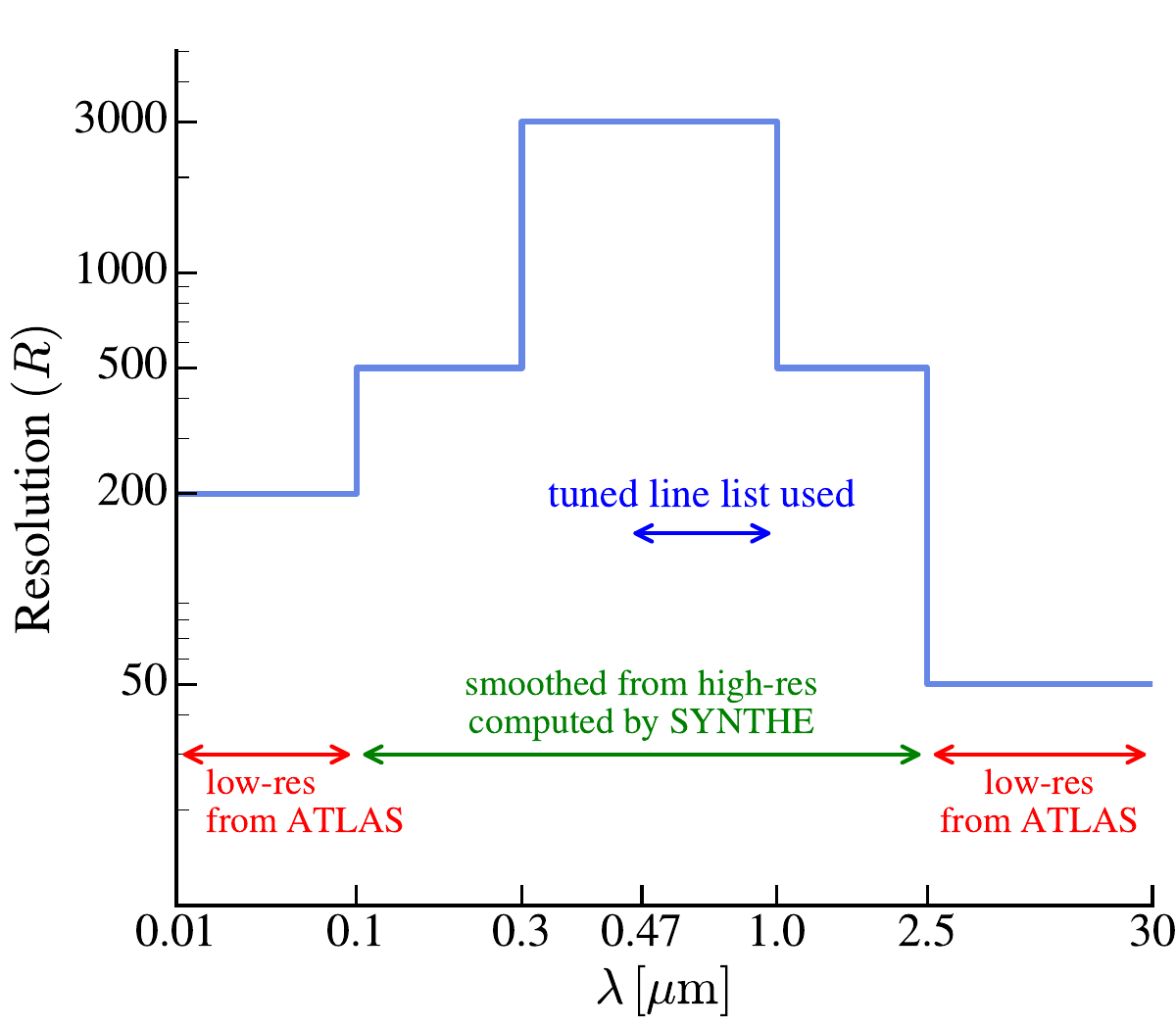}
    \caption{Resolution ($R=\rm \lambda/\Delta\lambda$) of the \newctk\ spectral library implemented in \texttt{FSPS} for different segments of the wavelength range. 
    A very high-resolution spectrum ($R\sim300,000$) for a wavelength range of $0.1-2.5\,\rm \mu m$ is computed by running a radiative transfer code, \texttt{SYNTHE}. 
    Outside this wavelength range ($100-1000\,\rm \AA$ and $2.5-30\,\rm \mu m$), a low-resolution SED computed by ATLAS is used.  
    The high-resolution spectrum from \texttt{SYNTHE} is then smoothed to different resolutions for more convenient use: $R=3000$ for $0.3-1.0\,\rm \mu m$. 
    The line list is tuned to the observed high-resolution spectra of the Sun and Arcturus for $\rm 4728\AA-1\,\rm \mu m$.} 
    \label{fig:c3k_resolution}
\end{figure}

The theoretical \ctk\ stellar spectra are computed from model atmospheres generated with \texttt{ATLAS12}, which assumes a 1D LTE plane-parallel atmosphere. 
For each star, a very low-resolution SED is computed for the wavelength range of $100\,\rm \AA -30\,\rm \mu m$. 
We then run a radiative transfer code, \texttt{SYNTHE} \citep[][]{Kurucz1970, Kurucz1981, Kurucz1993} through the model atmospheres to generate a very high-resolution ($R\sim300,000$) spectrum for each star over the wavelength range of $0.1-2.5\,\rm \mu m$.  
The latest atomic and molecular line list has been provided by R. Kurucz, and it has been tuned to the observed high-resolution spectra of the Sun and Arcturus for $\rm 4728\AA-1\,\rm \mu m$, as described in \cite{Cargile2020_C3K}. 
Finally, we smooth each stellar spectrum to different resolutions for different wavelength segments ($R=3000$ for $0.3-1.0\,\rm \mu m$) for more convenient use for SSPs. 
Fig.~\ref{fig:c3k_resolution} summarizes the resolutions used for different segments of the wavelength range and indicates which wavelength range is smoothed from the high-resolution spectrum computed by SYNTHE and which is from the low-resolution SED computed by ATLAS.
We adopt a microturbulence velocity of $v_{\rm micro}=1\,\rm km\,s^{-1}$, independent of spectral type.

The \ctk\ models cover a wide range of parameter space in \feh\ ($-2.5\leq\feh\leq+0.5$, in steps of 0.25), temperature ($2500\,{\rm K}\leq T_{\rm eff}\leq50,000\,{\rm K}$, approximately log-spaced), and gravity ($\rm -1.0\leq\log\,g\leq5.5$, in steps of 0.5).  
The models with varying $\alpha$-abundances (\afe=$-0.2,+0.0,+0.2,+0.4,+0.6$) are computed in lockstep by scaling up/down the abundance of all $\alpha$ elements (O, Ne, Mg, Si, S, Ar, Ca, and Ti) uniformly and simultaneously.

The key improvements of the \newctk\ models used in this work compared to the previous version presented in \cite{Byrne2022} are 1) the adopted solar abundances, 2) the use of a tuned line list, and 3) the updated microturbulence velocity from $v_{\rm micro}=2\,\rm km\,s^{-1}$ to $1\,\rm km\,s^{-1}$. 
The previous version of the \ctk\ models has used $v_{\rm micro}=2\,\rm km\,s^{-1}$. 
However, we have found a systematic discrepancy in the spectral indices between our models and the models using empirical libraries \citep{Knowles2023}. 
We found a better broad agreement in spectral indices when we used a lower microturbulence velocity. 
Thus, in the new version, we have adopted $v_{\rm micro}=1\,\rm km\,s^{-1}$, in agreement with observations showing that most stars have $v_{\rm micro}=1.0-1.5\,\rm km\,s^{-1}$ \cite[e.g.,][]{Bruntt2012}.
We adopt the solar abundances from \citetalias{GS98}, which is the same mixture used for the \newmist\ isochrones. 
Thus, the SSPs generated with \newmist\ and \newctk\ have fully consistent abundances.

\subsection{SSP construction}  \label{sec:model_ssp}
We use the \texttt{FSPS} framework \citep{Conroy2009_FSPS} to construct SSPs. 
To construct SSPs, the three main ingredients are used as follows: 
At a given age ($t$) and metallicity ([Fe/H] and \afe), the isochrone dictates which stars are to be included for the population synthesis, in terms of their mass, $T_{\rm eff}$ and $\log\,g$. 
The stellar spectral models contain stellar spectra for all these ($T_{\rm eff}$, $\log\,g$) grid points. 
The IMF assigns weights to stars of different masses ($m$). 
In short, SSPs are constructed by integrating the spectra of the stars along the isochrone, while taking into account the number of stars per mass bin following the adopted IMF. 
This can be written as the following equation: 
\begin{multline}
    F_\lambda(t, {\rm [Fe/H], [\alpha/Fe]}) \\= \int_{m_l}^{m_u} f_\lambda(t, m, {\rm [Fe/H], [\alpha/Fe]}) \phi(m) dm
\end{multline}
where $m_l$ and $m_u$ are the lowest and highest mass stars that are alive at a given age. 
In our models, $m_l=0.1\,M_\odot$, and $m_u(t)$ depends on age (determined by the isochrones).
The IMF is given by $\phi(m)$. 
Throughout this work, we use the Kroupa IMF \citep{Kroupa2001}, unless noted otherwise. 
The $f_\lambda$ is the stellar spectrum of a star at a given temperature and gravity.

We summarize the properties of our SSP models as follows.

\begin{itemize}
\setlength\itemsep{0.0em} 
    \item Ages: $\log(\rm age/yr)=5.0-10.3$ in steps of 0.05 dex
    \item Wavelength range: $100\,\rm \AA-30\,\rm \mu m$
    \vspace{-0.1cm}
    \begin{itemize}[leftmargin=*]
    \setlength\itemsep{-0.2em} 
        \item High-resolution from \texttt{SYNTHE} for $0.1-2.5\,\rm \mu m$
        \item Smoothed to $R=3000$ for $0.3-1.0\,\rm \mu m$
        \item Smoothed to a lower resolution elsewhere
    \end{itemize}
    \vspace{-0.1cm}
    \item \feh: $[-2.5, +0.5]$ in steps of 0.25
    \item \afe: $-0.2, +0.0, +0.2, +0.4, +0.6$
    \item Solar mixture: \cite{GS98} \\ (adopted for both isochrones and stellar spectra)
\end{itemize}

\subsection{Caveats \& Limitations}
\label{sec:caveats_limitations}

We describe the caveats \& limitations of our models. 
The \feh\ and \afe\ refer to the initial composition of stars, and changes in surface abundances as a result of the stellar evolutionary processes (e.g., diffusion and dredge-up) are not reflected in our stellar atmosphere calculations.  
All $\alpha$ elements are locked into a single value of \afe, instead of varying individual $\alpha$ abundances. 
The C/N variations have not been taken into account, and their surface abundances are kept to be solar-scaled throughout all stellar evolutionary phases. 
The \ctk\ theoretical spectra libraries are computed based on the 1D plane-parallel model atmospheres assuming LTE. 
While the microturbulence velocity varies with stellar temperatures and gravity \citep[e.g.,][]{Bruntt2012}, we have adopted a constant microturbulence velocity ($v_{\rm micro}=1\,\rm km\,s^{-1}$) for all stellar types.

\section{Results}
\subsection{Comparison of \oldmist\ to \newmist\ isochrones at solar \afe} \label{sec:result_compare_old_mist_new_mist}

\begin{figure*}[!]
\centering
\renewcommand{\arraystretch}{0}
\setlength{\tabcolsep}{0pt}
\begin{tabular}{cc}
\includegraphics[width=1.05\columnwidth]{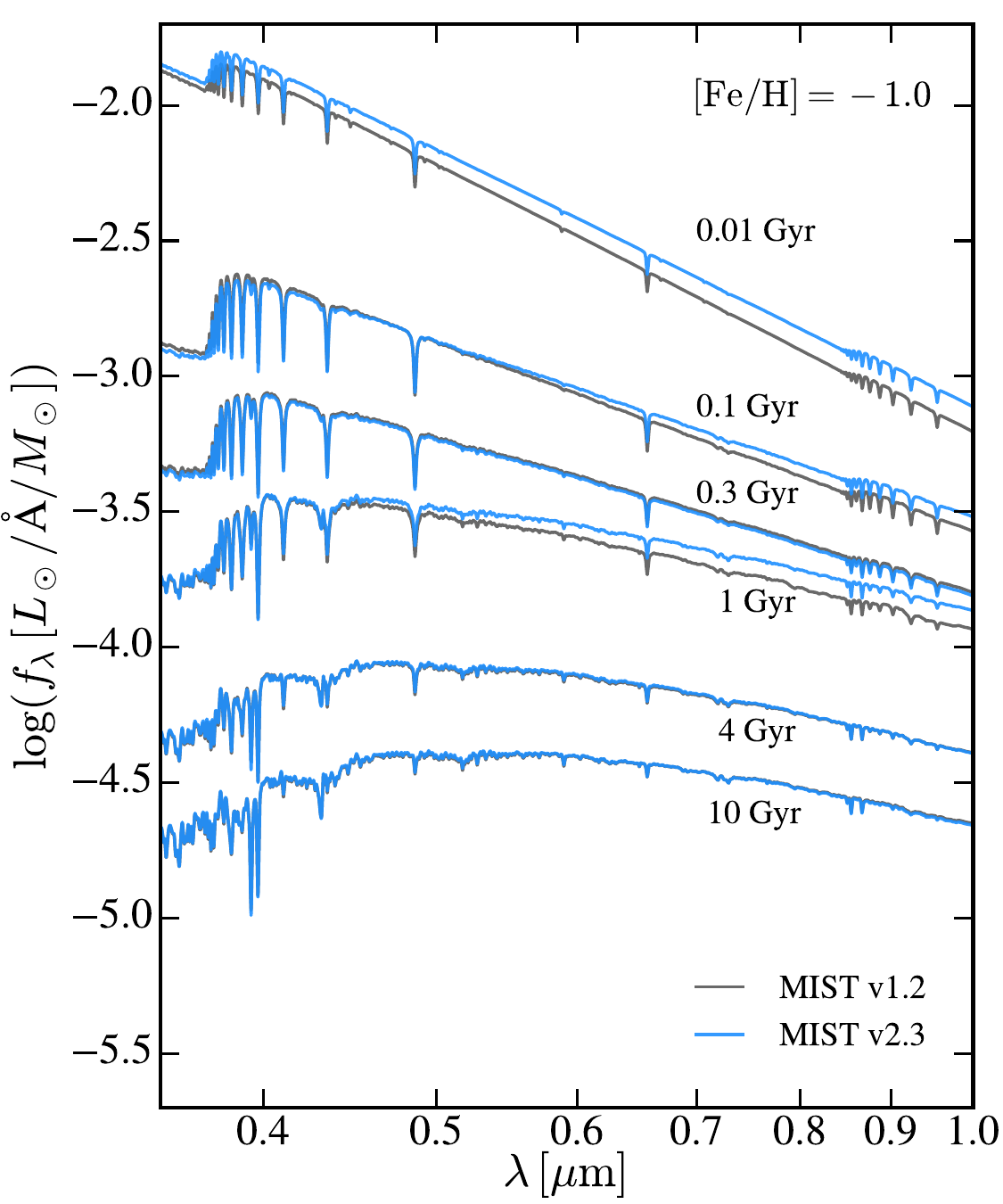} &
\includegraphics[width=1.05\columnwidth]{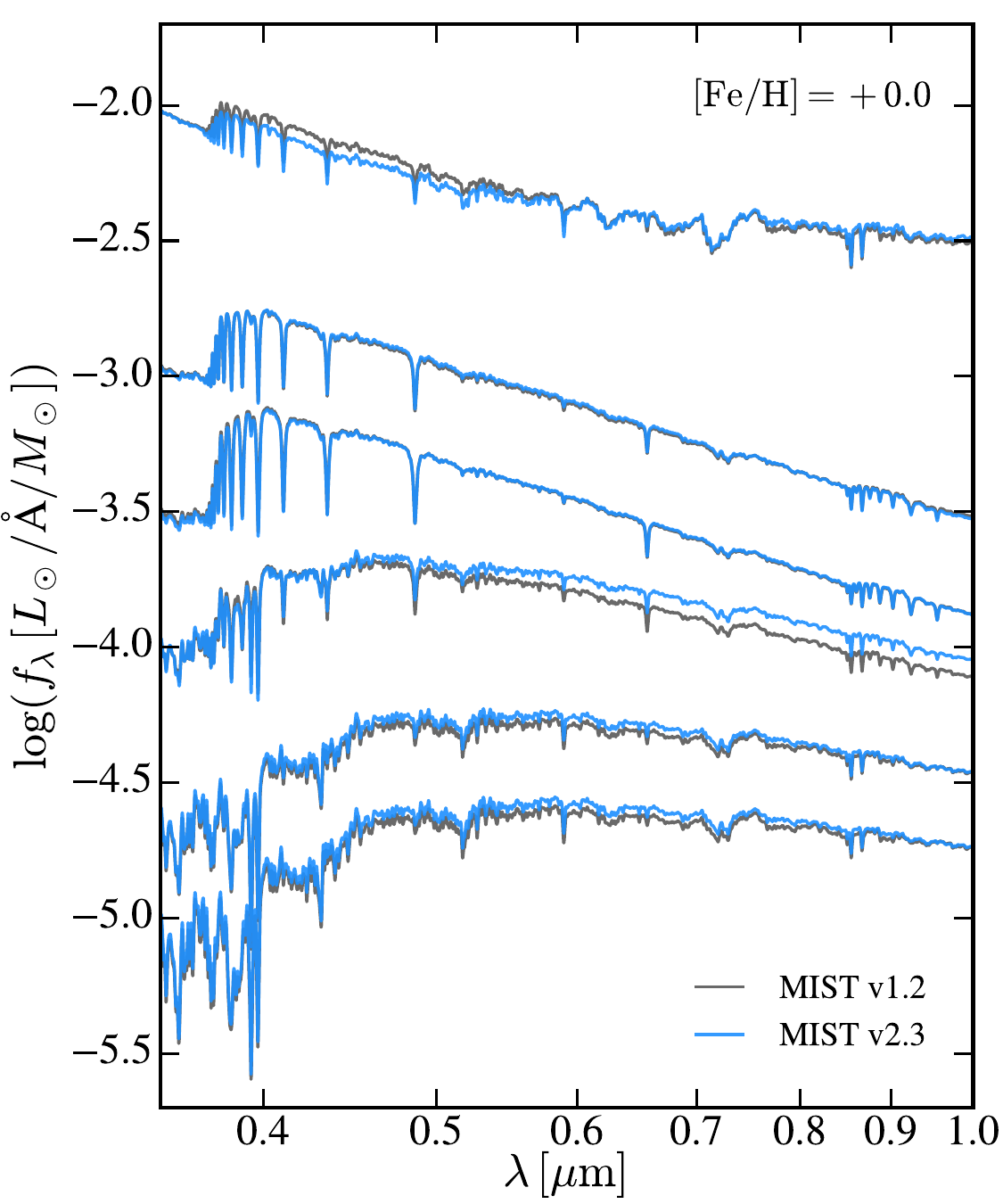}
\end{tabular}
\caption{Comparison of SSP full spectra generated using the new \newmist\ (blue, used in this work) and previous \oldmist\ (gray) for different ages. We find a good overall agreement between SSP spectra made with \newmist\ and \texttt{v1.2}.}
\label{fig:sed_mist1_mist2}

\vspace{0.05cm}

\end{figure*}
\begin{figure*}[!]
    \centering
    \includegraphics[width=0.97\textwidth]{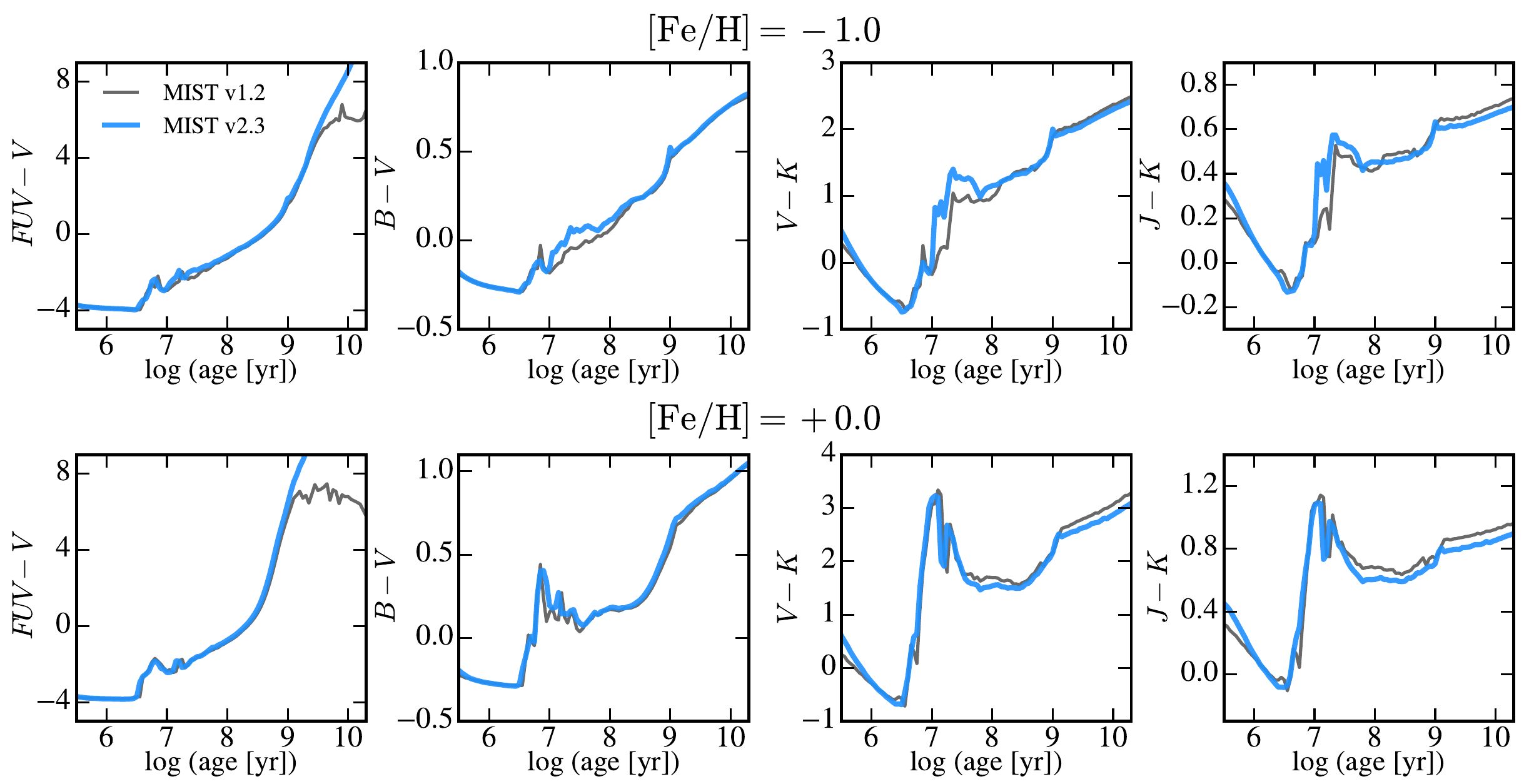}
    \caption{
    Comparison of the color evolutions of SSP generated with \oldmist\ \citep{Choi2016} (gray) and the new \newmist\ (this work, blue) for \feh$=-1.0$ (top) and \feh$=+0.0$ (bottom) at \afe$=+0.0$. 
    We find good overall agreement in SSP colors between \oldmist\ and \newmist. 
    One noticeable difference is found in $FUV-V$ color for SSPs with ages $\rm \log(age/yr)>9.0$. 
    This is because the post-AGB and WD phases are included in the previous \oldmist\ models, but not in the new \newmist\ models.}
    \label{fig:color_mist1_mist2}
\end{figure*}

First, we compare our SSP models generated with the new \newmist\ isochrones to those generated with previous \oldmist\ presented in \cite{Choi2016} at solar $\alpha$-abundance (\afe$=+0.0$). 
The new \newctk\ is used for both cases. 
Therefore, the major difference should be the base solar mixture: \cite{Asplund2009} in v1.2 and \citetalias{GS98} in the new \newmist.  
Fig.~\ref{fig:sed_mist1_mist2} compares the SSP spectra for 0.01, 0.1, 0.3, 1, 4, and 10 Gyr at \afe$=+0.0$ for \feh$=-1.0$ (left) and \feh$=+0.0$ (right); the blue lines show the SSPs with the new \newmist, while the gray lines are based on \oldmist.
Overall, the continuum shapes look consistent between the previous and new \mist\ models. 
For the detailed analysis of the continuum shapes, we compare the broadband colors of SSPs.

Fig.~\ref{fig:color_mist1_mist2} compares the time evolution of colors ($FUV-V$, $B-V$, $V-K$, and $J-K$) for \feh$=-1.0$ (top) and \feh$=+0.0$ (bottom). 
The same color code is used for the previous and new \mist\ models. 
We find that the previous and new \mist\ models are in good agreement with the SSP colors. 
The large spikes toward red colors in $V-K$ and $J-K$ at $\rm \log(age/yr)\sim7.0$ are due to the appearance of red supergiants (RSG), and the onset of this RSG phase agrees well between the previous and new models. 
There is a noticeable difference in $FUV-V$ colors where SSPs generated with \oldmist\ turn over toward bluer colors for $\rm \log(age/yr)>9.0$, while SSPs generated with \newmist\ continue to become redder. 
This is because the post-AGB and white dwarf (WD) phases are included in \oldmist, but not in \newmist. 

\subsection{The effect of $\alpha$-enhancement on SSPs}  \label{sec:result_afsps_compare_solar_vs_afe}
In this section we investigate the effect of varying \afe\ on SSPs at fixed \feh. 
We explore three SSP outputs: 1) full spectra, 2) broadband colors, and 3) spectral indices. 


\subsubsection{SSP Spectra}

\begin{figure*}
    \centering
    \includegraphics[width=1\textwidth]{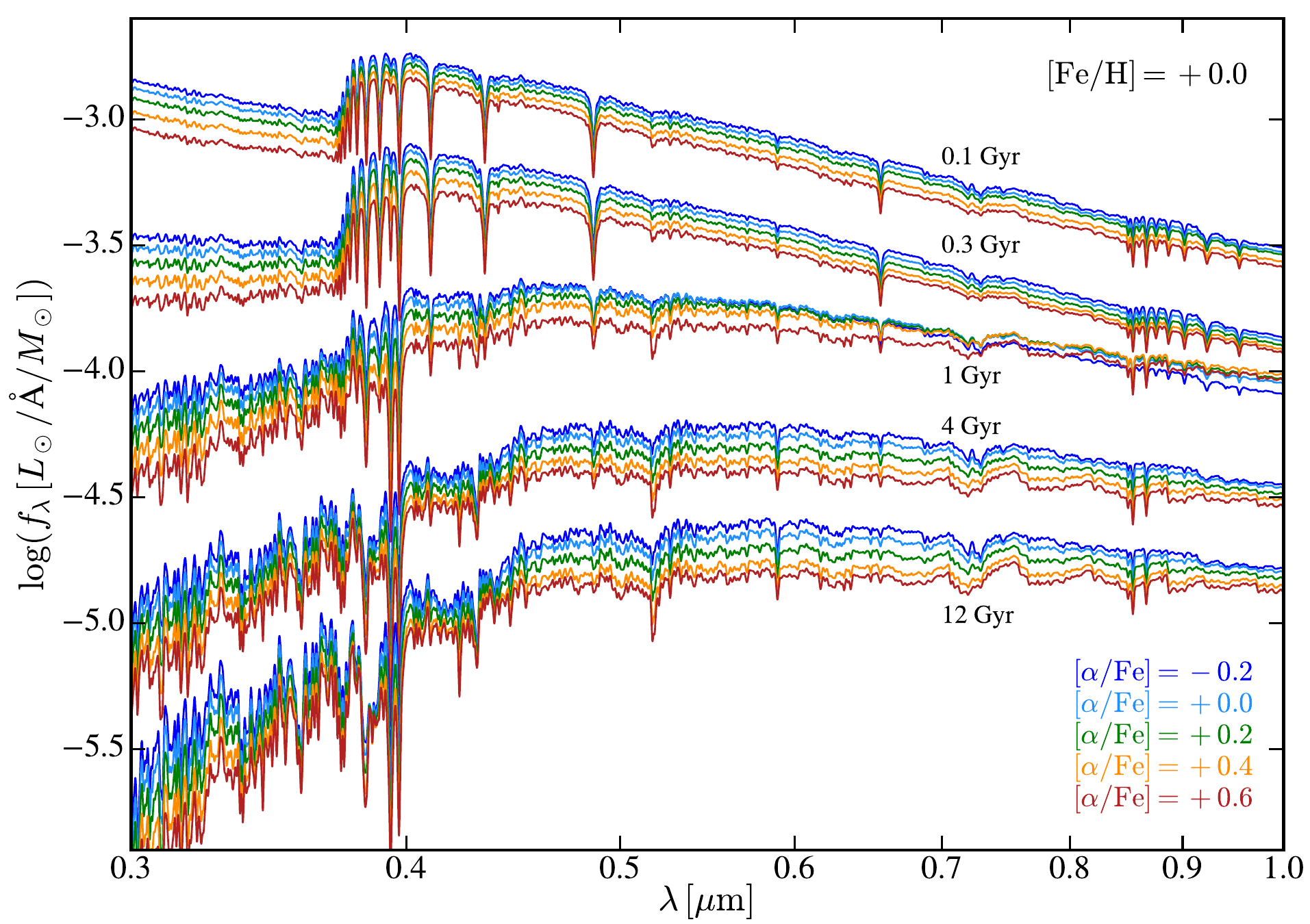}
    \caption{SSP spectra with ages of 0.1, 0.3, 1.0, 4.0, and 12 Gyr with fixed \feh$=+0.0$ and varying \afe\ (shown as different colors).} 
    \label{fig:afsps_spectra}
\end{figure*}

\begin{figure*}[hbt!]
    \centering
    \includegraphics[width=1.0\textwidth]{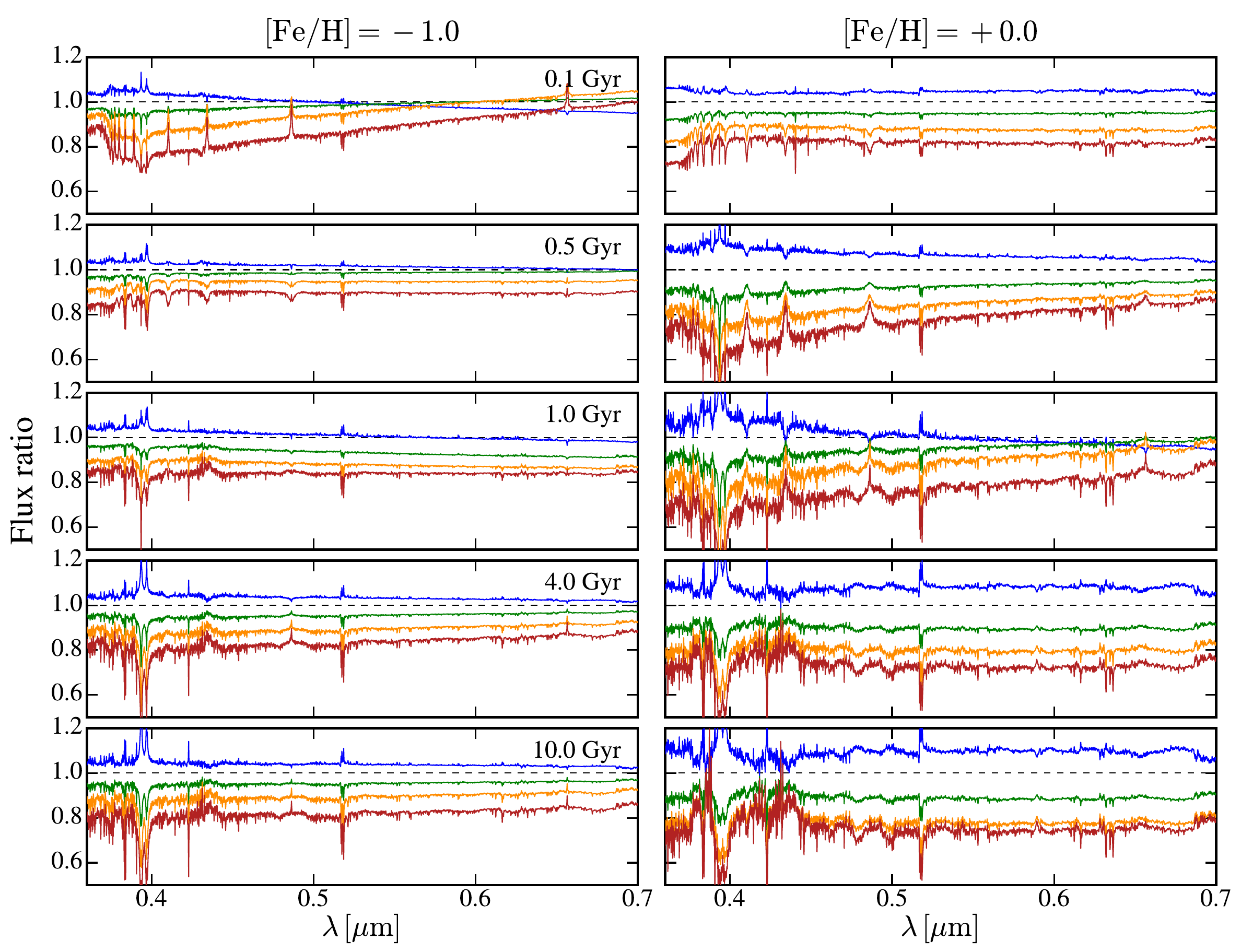}
    \caption{
    Flux ratios between $\alpha$-enhanced and solar-scaled SSPs at fixed \feh: \feh$=-1.0$ (left) and $+0.0$ (right). 
    From top to bottom, we show the flux ratios of SSPs for different ages, from 0.1 Gyr to 10 Gyr. 
    For fixed \feh, different color represents different \afe: $-0.2$ (blue), $+0.2$ (green), $+0.4$ (orange), $+0.6$ (brown). } 
    \label{fig:afsps_flux_ratios}
\end{figure*}

\begin{figure*}[hbt!]
    \centering
    \includegraphics[width=1.0\textwidth]{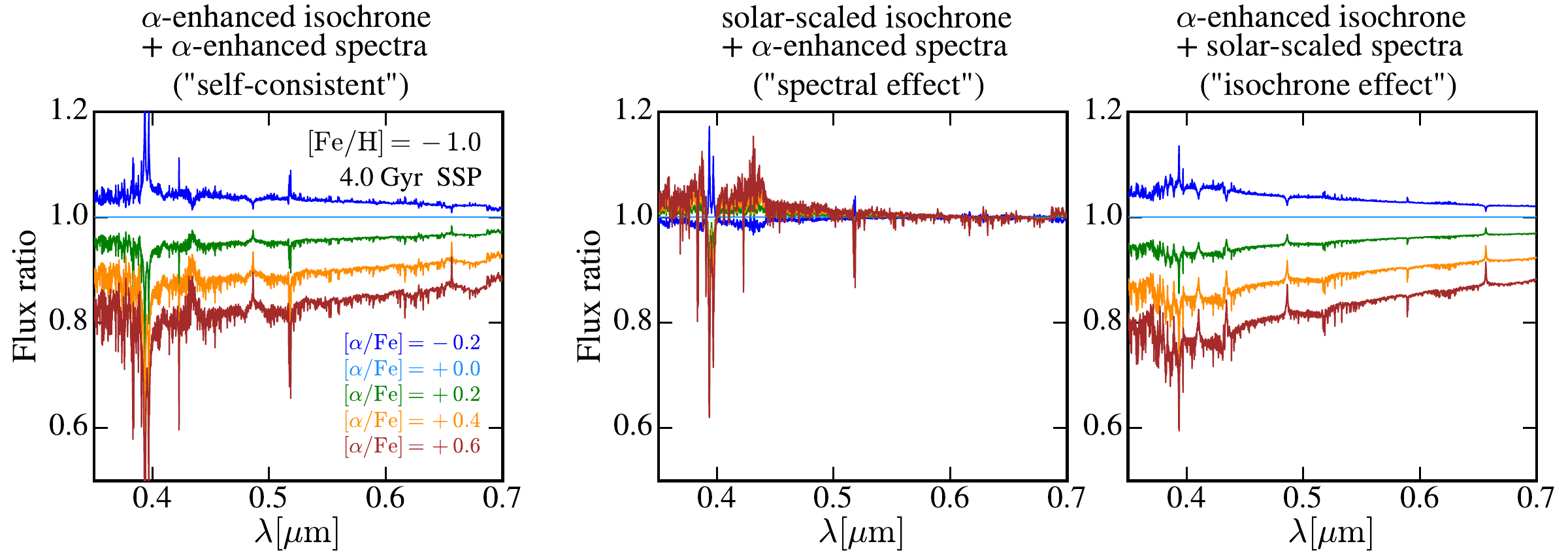}
    \caption{Respective $\alpha$-enhancement effects of isochrones and stellar spectra on SSP spectra. 
    All three panels show the flux ratios between $\alpha$-enhanced and solar-scaled SSPs at fixed \feh$=-1.0$ for 4 Gyr. The same color code is used for different \afe\ ratios. 
    The $\alpha$-enhanced models in each panel are produced as follows:
    (Left) $\alpha$-enhanced isochrones and $\alpha$-enhanced stellar spectra, the ``self-consistent'' $\alpha$-enhanced models. 
    (Middle) Solar-scaled isochrones and $\alpha$-enhanced stellar spectra, showing the ``spectral effect''.  
    (Right) $\alpha$-enhanced isochrones and solar-scaled stellar spectra, showing the ``isochrone effect''. 
    We find that the main effect of $\alpha$-enhancement in isochrones is to lower the overall continuum and change the continuum shapes, whereas $\alpha$-enhancement in stellar spectra mainly affects the individual spectral lines. } 
    \label{fig:afsps_flux_ratios_solariso_solarc3k}
\end{figure*}

Fig.~\ref{fig:afsps_spectra} shows SSP spectra with ages of 0.1, 0.3, 1, 4, and 12 Gyr for \feh$=+0.0$ and varying \afe\ (shown as different colors). 
To explore the effect of $\alpha$-enhancement in more detail, we take the flux ratios between $\alpha$-enhanced and solar-scaled SSPs. 
Fig.~\ref{fig:afsps_flux_ratios} shows the flux ratios between $\alpha$-enhanced and solar-scaled SSPs: \feh$=-1.0$ (left) and $+0.0$ (right). 
The same color codes are used as in Fig.~\ref{fig:afsps_spectra}. 
All SSPs are smoothed to a resolution of $\sigma=250\,\rm km\,s^{-1}$.

At fixed \feh, one of the main effects of $\alpha$-enhancement is lowering the overall continuum level. 
The average flux ratio for \afe$=+0.4$ model (at \feh$=+0.0$) is $\sim0.8$. 
This is due to the increased $\rm H^{-}$ opacity. 
The $\alpha$ elements, especially Mg, are important electron donors, contributing to increasing the $\rm H^{-}$ opacity. 
The effect of lowering continuum is more pronounced for higher \feh\ and older SSP ages, as the continuum of G, K, and  M-type stars is mainly formed by free-free and bound-free transitions of $\rm H^{-}$.  
The detailed continuum shapes and resulting broadband colors will be presented in the following section. 
In addition to the overall continuum level, spectral lines sensitive to $\alpha$-elements are deeper in the $\alpha$-enhanced models due to increased abundances, such as Ca II H\&K ($\sim 3950\,\rm \AA$), and Ca I line ($\sim 4227\,\rm \AA$), and Mg triplet ($\sim 5175\,\rm \AA$). 
Higher-order Balmer lines also appear to be sensitive to $\alpha$-enhancement.
A more detailed analysis of the effect of $\alpha$-enhancement on spectral indices will be presented in Section~\ref{sec:results_spectral_indices}.

We separate the $\alpha$-enhancement effects of isochrones and stellar spectra on the overall SSPs. 
In Fig.~\ref{fig:afsps_flux_ratios_solariso_solarc3k}, all three panels show the flux ratios between the $\alpha$-enhanced and solar-scaled SSPs with ages of 4 Gyr and \feh$=-1.0$. 
The left panel shows the $\alpha$-enhanced models where both isochrones and stellar spectra are $\alpha$-enhanced (``self-consistent''). 
The middle panel shows the ``spectral effect'' of $\alpha$-enhancement with models where only stellar spectra are $\alpha$-enhanced. 
The right panel shows the ``isochrone effect'' where only isochrones are $\alpha$-enhanced.

We find that the primary effect of $\alpha$-enhancement in isochrones is to lower the overall continuum and change the continuum shapes, whereas, $\alpha$-enhancement of stellar spectra mainly affects the individual spectral lines. 
In the middle panel showing the ``spectral effect'', the average flux ratio to the solar-scaled models is almost one. 
However, individual spectral lines, especially, the lines sensitive to $\alpha$ abundances, e.g., Ca and Mg lines show great sensitivity to $\alpha$-enhancement in stellar spectra. 
While the spectral effect seems to be more dominant in individual spectral lines, $\alpha$-enhancement in isochrones also makes most of the lines deeper, except for the Balmer lines which are weaker in the $\alpha$-enhanced models (appearing as upward spikes in the right panel). 
A more detailed analysis of separating the isochrone and spectral effects on individual spectral indices will be presented in Section~\ref{sec:results_spectral_indices} (See Fig.~\ref{fig:afsps_lick_indices_solariso_solarc3k}).

\subsubsection{Colors}
\label{sec:results_colors}

\begin{figure*}[hbt!]
    \centering
    \includegraphics[width=1.0\textwidth]{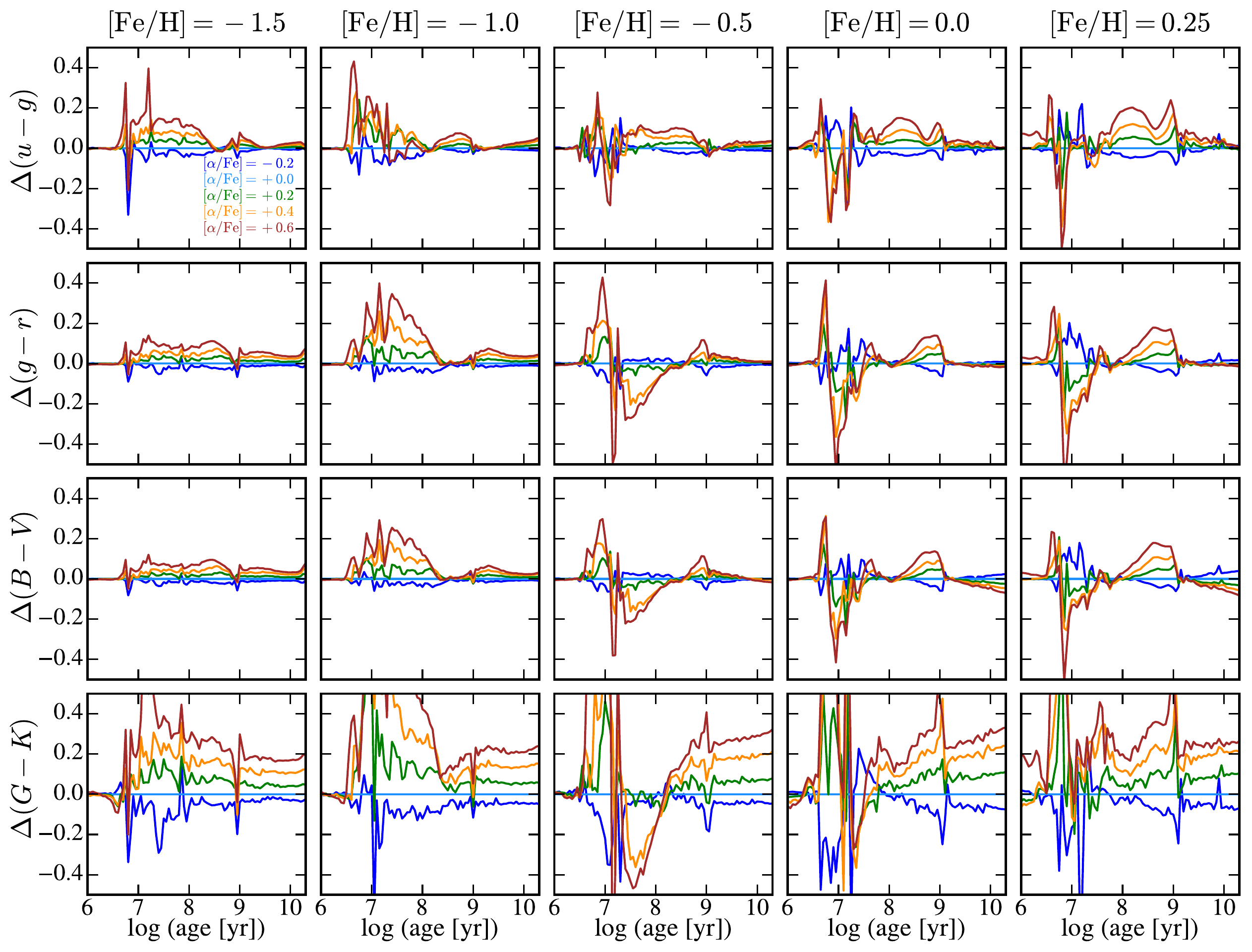}
    \caption{Differences in broadband SSP colors ($u-g$, $g-r$, $B-V$, and $G-K$) between $\alpha$-enhanced models (different colors for varying \afe) and solar-scaled models at constant \feh\ (each \feh\ shown in each column). For older stellar population ($>1\,\rm Gyr$), we find that $\alpha$-enhanced models have redder $G-K$ color (by $\sim0.2$ mag for \afe$=+0.4$) and do not show significant differences in colors at shorter wavelengths. The $\alpha$-enhancement has significant impacts on the colors for young stellar populations ($\rm \log(age/yr)=7-9$).}
    \label{fig:afsps_color_evolution}
\end{figure*}

\begin{figure}[hbt!]
    \centering
    \includegraphics[width=1.0\columnwidth]{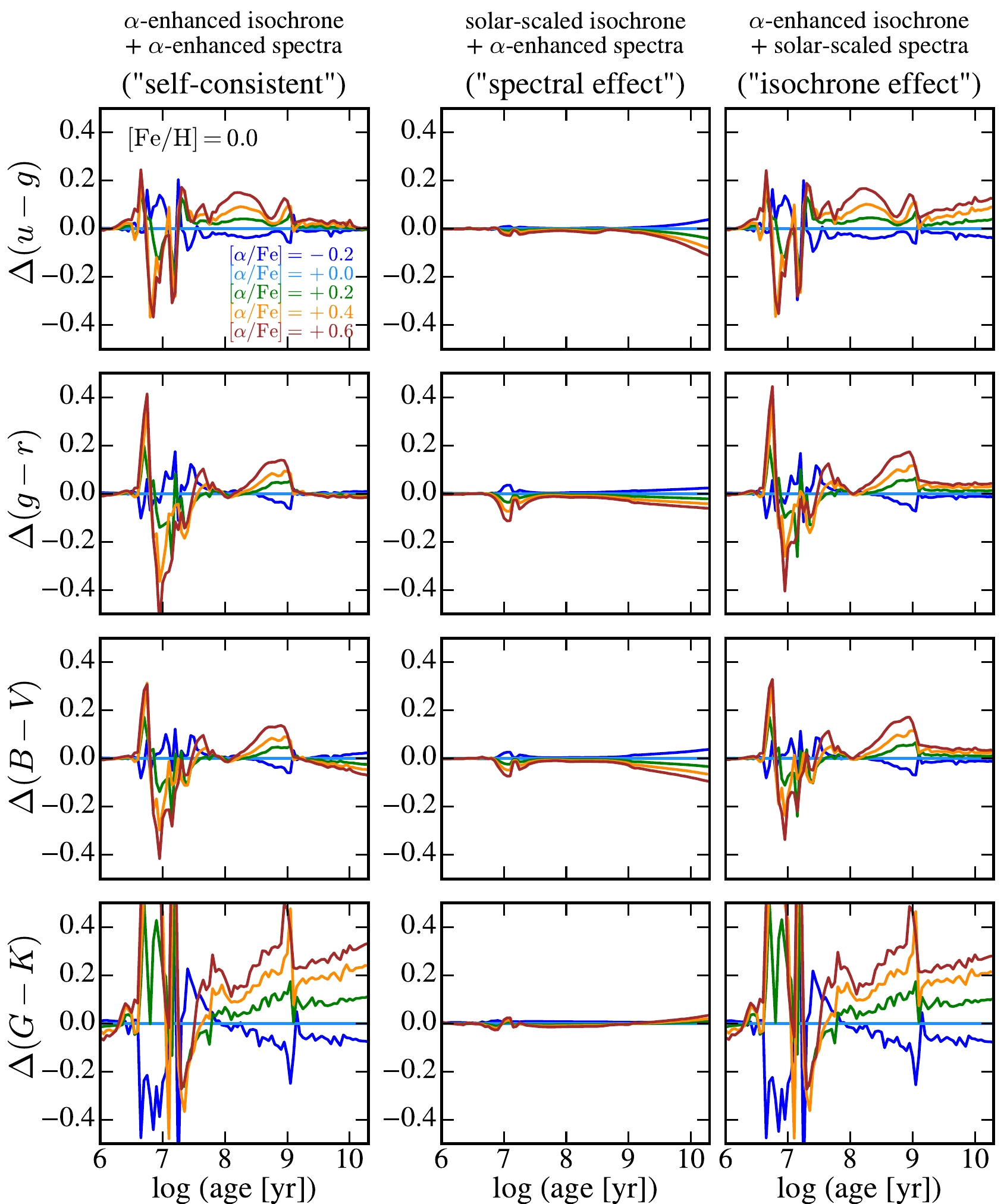}
    \caption{Separate effects of $\alpha$-enhancement in isochrones vs. stellar spectra on SSP colors. All panels show the color differences between $\alpha$-enhanced and solar-scaled models at \feh$=+0.0$ (same format as in Fig.~\ref{fig:afsps_color_evolution}). The $\alpha$-enhanced models in each column are as follows:
    (Left) ``self-consistent'' $\alpha$-enhanced models where both isochrones and stellar spectra are $\alpha$-enhanced. (Middle) Only stellar spectra are $\alpha$-enhanced, showing the ``spectral'' effects on the broadband colors. (Right) Only isochrones are $\alpha$-enhanced, showing the ``isochrone'' effects. We find that $\alpha$-enhancement in isochrones and stellar spectra have opposite effects on broadband colors for older stellar populations ($>1\,\rm Gyr$). $\alpha$-enhancement in isochrones always makes SSPs redder in all colors by lowering the temperature of stars dominating the fluxes. On the other hand, $\alpha$-enhancement in stellar spectra makes them bluer in bluer bands ($u-g$, $g-r$, and $B-V$).} 
    \label{fig:afsps_color_solariso_solarc3k}
\end{figure}

In this section we study the effects of $\alpha$-enhancement on broadband colors ($u-g$, $g-r$, $B-V$, and $G-K$) at fixed \feh.
Fig.~\ref{fig:afsps_color_evolution} shows the color differences between $\alpha$-enhanced (lines with different colors for varying \afe) and solar-scaled SSPs at constant \feh\ (shown in each column). 
We adopt the AB magnitude system for the $u$, $g$, and $r$-band filters defined by the Sloan Digital Sky Survey (SDSS) filter profiles. 
The Vega magnitude system is adopted for the Johnson $B$, $V$ bands, 2MASS-$Ks$ band, and {\it Gaia G} band, where we use the Vega spectrum from \cite{Bohlin2007, Bohlin2014}.

For old stellar populations ($\rm >1 Gyr$), the $G-K$ color is redder in the $\alpha$-enhanced models at constant \feh, due to increased total metallicity.  
For \feh$=+0.0$, $\alpha$-enhanced models with \afe$=+0.4$ have redder $G-K$ color by $\sim0.2$ mag (Fig.~\ref{fig:afsps_color_evolution}) for all ages. 
This is consistent with the results of other previous studies. 
For example, \cite{Coelho2007} and \cite{Percival2009} showed that $\alpha$-enhanced models at constant \feh\ have redder colors in the red continuum (for $V-I$, $R-I$, $g-i$, and $g-z$ colors).  
The reddening at longer wavelengths is mainly due to the increased total metallicity. 
We will discuss the effect of $\alpha$-enhancement on broadband colors at fixed \zh\ in Section~\ref{sec:results_fixz}.

To see the individual $\alpha$-enhanced effects of isochrones and spectra on broadband colors more clearly, we separate these effects in Fig.~\ref{fig:afsps_color_solariso_solarc3k}. 
All panels show the color differences between $\alpha$-enhanced and solar-scaled models at \feh$=+0.0$ (same format as in Fig.~\ref{fig:afsps_color_evolution}). 
The left panel shows the differences in color between ``self-consistent'' $\alpha$-enhanced models (i.e., both isochrones and stellar spectra are $\alpha$-enhanced) and solar-scaled models at \feh$=0.0$. 
In the middle panel, only stellar spectra are $\alpha$-enhanced in the $\alpha$-enhanced models, showing the ``spectral'' effects on the broadband colors. 
In contrast, the right panel shows the ``isochrone'' effects where only isochrones are $\alpha$-enhanced.

For old ages ($\rm >1 Gyr$), the effect of $\alpha$-enhancement (at fixed \feh) on colors at shorter wavelengths seems to be more subtle as a result of the opposite effects of isochrones and stellar spectra.  
Fig.~\ref{fig:afsps_color_solariso_solarc3k} shows that $\alpha$-enhancement in isochrones always makes SSPs redder in all colors by lowering the temperature of stars dominating the fluxes. 
On the other hand, $\alpha$-enhancement in stellar spectra makes SSPs bluer at shorter wavelengths ($u-g$, $g-r$, and $B-V$), especially for older ages.  
Several previous studies have also found that $\alpha$-enhancement in stellar spectra increases the fluxes at bluer wavelengths. 
For example, \cite{Cassisi2004} has found that $\alpha$-enhancement makes stellar colors bluer in $U-B$ and $B-V$, and the difference is larger at higher metallicity and lower $T_{\rm eff}$. 
They further investigated the contributions of individual $\alpha$-abundances to this difference and found that Mg is the most responsible for affecting the continuum flux distributions. 
\cite{Coelho2007} has also shown that bluer $U-B$ in the $\alpha$-enhanced models (by $\sim 0.1$ mag) is due to the spectral effect. 
\cite{Choi2019} explored the colors of $\alpha$-enhanced models using \alf\ \citep{Conroy2018_alf} (where only stellar spectra are $\alpha$-enhanced) and found much bluer $u-g$ and $g-r$ colors in the $\alpha$-enhanced models due to the increased fluxes at blue wavelengths.

$\alpha$-enhancement also has significant impacts on the broadband colors for younger stellar populations ($<\rm 1\,Gyr$). 
The SSP colors for ages between $\rm \log(age/yr)=7-8$ are associated with the onset of the RSG phases of massive stars. 
As mentioned above, the large spike at $\rm \log(age/yr)=7-8$ toward red colors in $G-K$ is due to the appearance of RSGs.
In the $\alpha$-enhanced models, the onset of the RSG phase seems to appear at earlier times than in solar-scaled models. 
The reddening with $\alpha$-enhancement at ages between $\rm \log(age/yr)=8-9$ is associated with the TP-AGB phases. 
As the effect of $\alpha$-enhancement on colors for younger stellar populations is related to stellar evolutionary phases, it is only shown with the isochrone effects in Fig.~\ref{fig:afsps_color_solariso_solarc3k}.
A more detailed study of the effect of $\alpha$-enhancement on these specific stellar evolutionary phases will be presented in Dotter et al. in prep.

\subsubsection{Spectral indices}
\label{sec:results_spectral_indices}

\begin{figure*}[hbt!]
    \centering
    \includegraphics[width=1.0\textwidth]{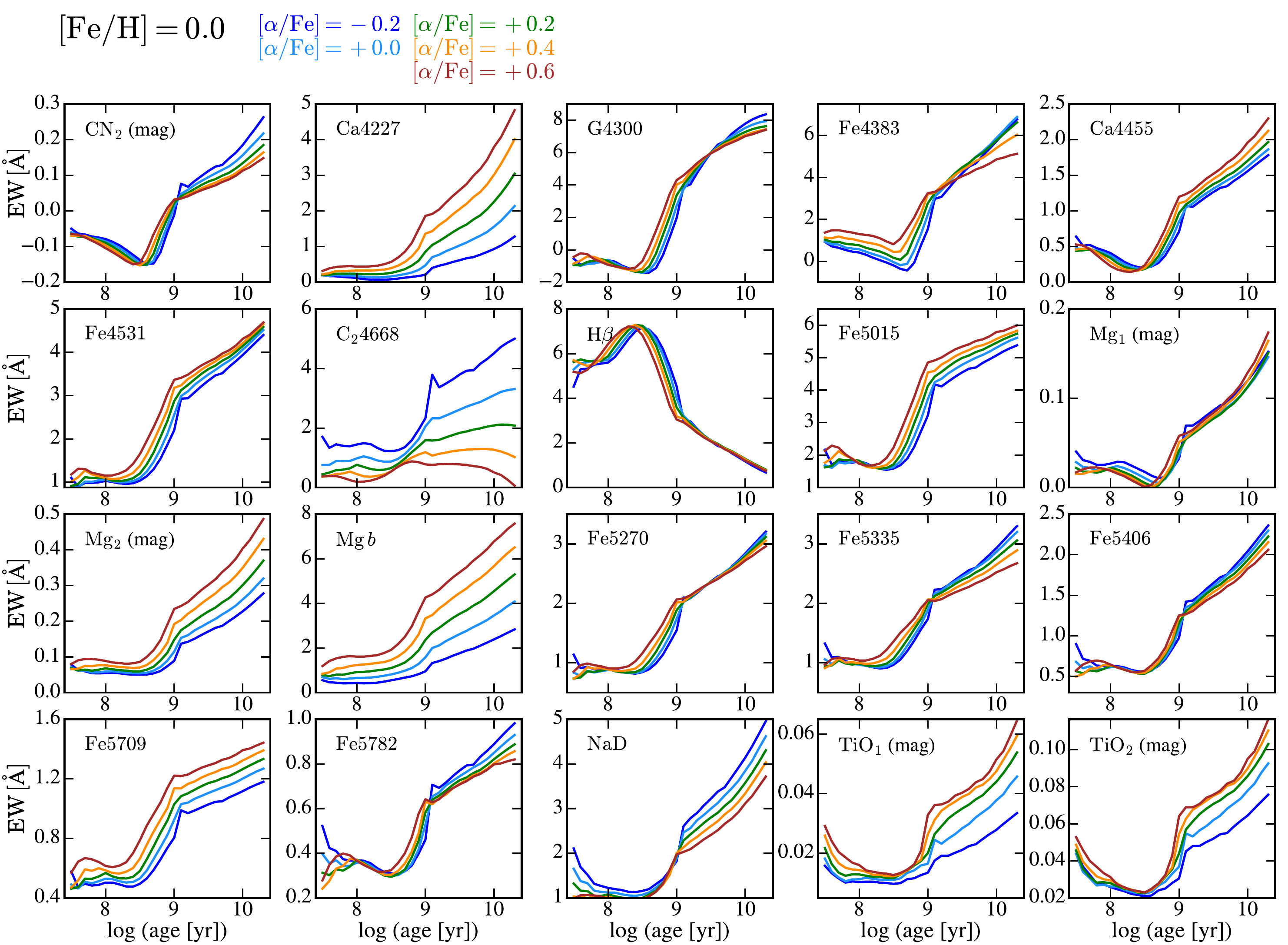}
    \caption{Lick indices as a function of SSP ages. Different colors represent SSPs with different \afe\ at constant \feh$=+0.0$. All indices are measured from SSP spectra at a resolution of $8.4\,\rm \AA$ (FWMH). We find that spectral indices involving $\alpha$-elements (e.g., Ca4227, \mgb, $\rm TiO$) increase with \afe, while most iron- and carbon-sensitive indices get weakened with increasing \afe. }
    \label{fig:afsps_lick_indices}
\end{figure*}

\begin{figure*}[hbt!]
    \centering
    \includegraphics[width=1.0\textwidth]{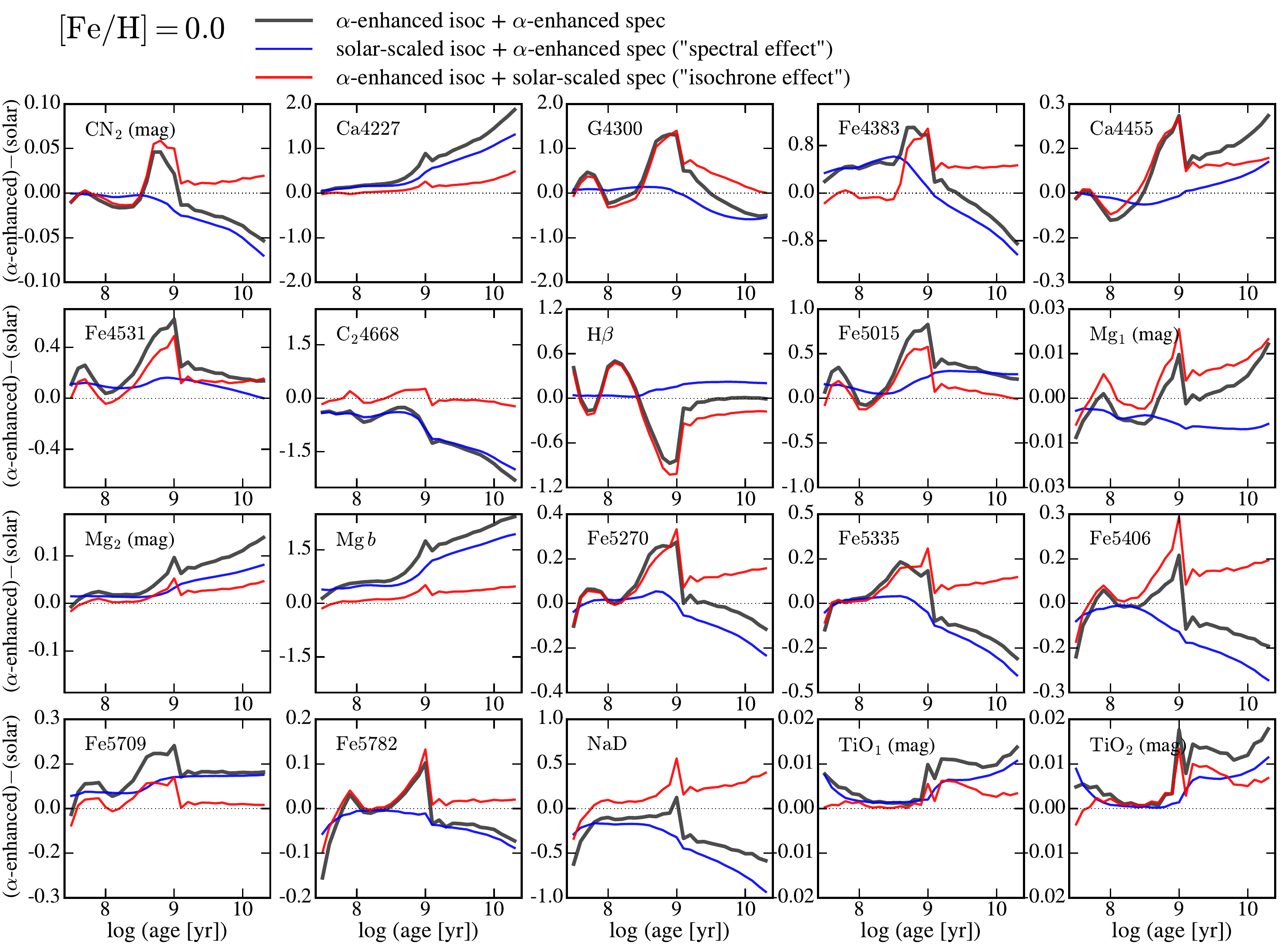}
    \caption{Differences in spectral indices between the $\alpha$-enhanced (\afe$=+0.4$) and solar-scaled SSPs at fixed \feh$=+0.0$ as a function of ages. The thick black lines show the differences from the $\alpha$-enhanced models where both isochrones and stellar spectra are $\alpha$-enhanced. The blue and red lines show the ``spectral effect'' and ``isochrone effect'', respectively, by adopting $\alpha$-enhanced models only for stellar spectra and isochrones, respectively. In most cases, we find the opposite effects of isochrones and stellar spectra on spectral indices. $\alpha$-enhancement in isochrones increases the line strengths by lowering the temperature of stars dominating the SSP fluxes. On the other hand, $\alpha$-enhancement in stellar spectra makes most indices weaker, except for the indices involving $\alpha$-elements where the enhanced $\alpha$-abundances make them stronger.}
    \label{fig:afsps_lick_indices_solariso_solarc3k}
\end{figure*}

\begin{figure*}[hbt!]
    \centering
    \includegraphics[width=1.0\textwidth]{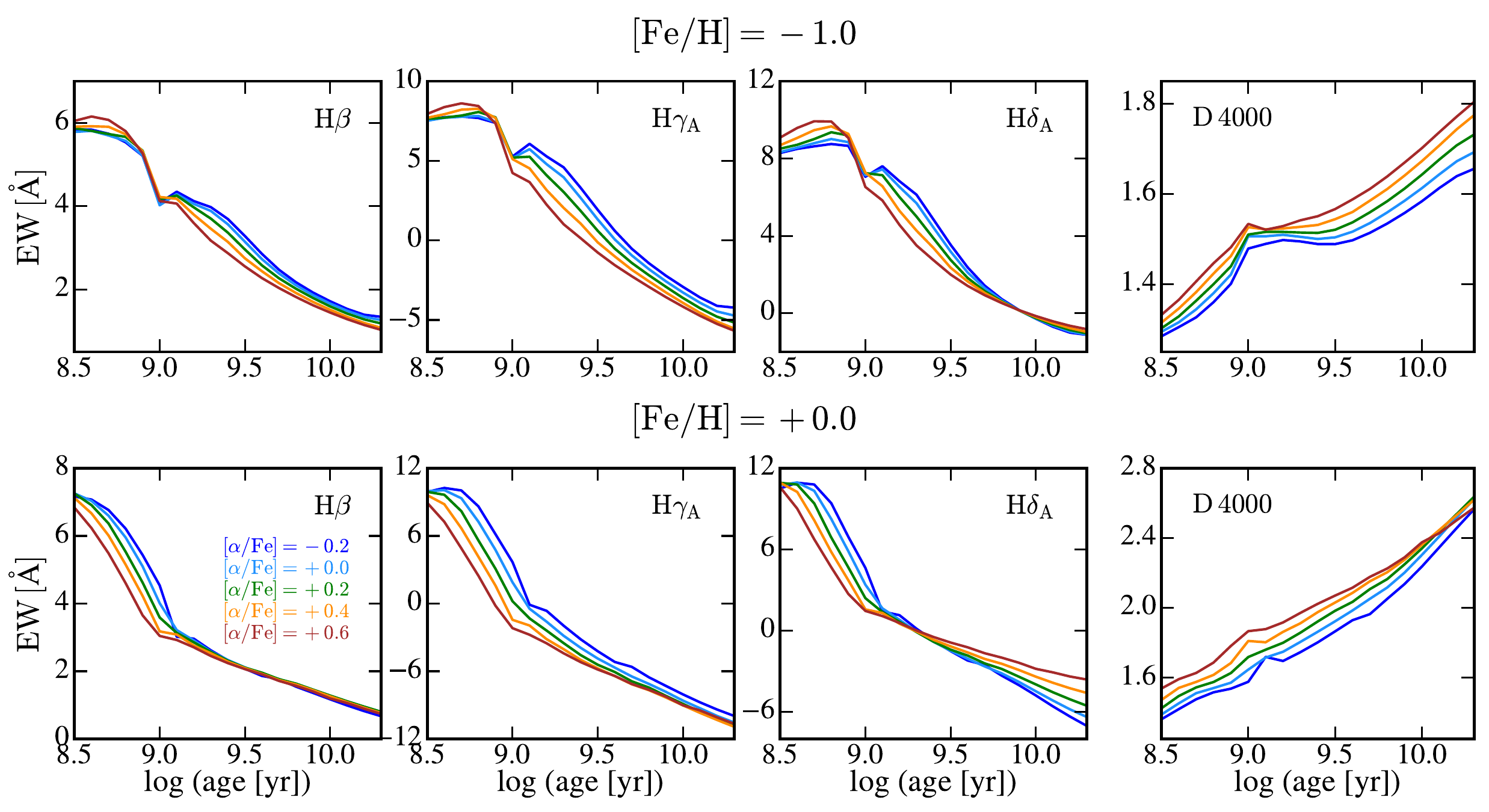}
    \caption{Age-sensitive indices ($\rm H\beta$, $\rm H\gamma_A$, $\rm H\delta_A$, and D4000) as a function of ages (zooming for $\rm \log(age/yr)=8.5-10.3$) for SSPs with \feh$=-1.0$ (top) and \feh$=+0.0$ (bottom). The same color code is used for different \afe\ as in Fig.~\ref{fig:afsps_lick_indices}. We find that the $\alpha$-enhanced models at fixed \feh\ show weaker Balmer strengths and the SSP ages the most sensitive to \afe\ depend on \feh. 
    Therefore, to accurately derive the ages of stellar populations, a detailed analysis with chemical abundances (\feh,\afe) is required. 
    }
    \label{fig:afsps_balmer_lines}
\end{figure*}

In this section we study the effect of $\alpha$-enhancement (at fixed \feh) on individual spectral indices.
We measure the Lick indices, defined in \cite{Trager2000}, from SSP spectra at a resolution of $8.4\,\rm \AA$ (FWHM), assuming the Kroupa IMF \citep{Kroupa2001}. 
Fig.~\ref{fig:afsps_lick_indices} shows the Lick indices as a function of SSP ages at constant \feh$=+0.0$ with varying \afe\ (shown as different colors). 
We also show SSP spectra of a fixed age (4 Gyr) zooming in around individual spectral features in Fig.~\ref{fig:afsps_individual_lines}, highlighting the index bandpass and the blue and red pseudo-continua defined in \cite{Trager2000}. 
We emphasize again that we explore the effect of $\alpha$-enhancement at constant \feh\ (not at constant total metallicity) so that our $\alpha$-enhanced models have higher total metallicities than solar-scaled models.

We further separate the isochrone and spectral effects on individual line indices in Fig.~\ref{fig:afsps_lick_indices_solariso_solarc3k}. 
Each panel shows the difference in each index measured from the $\alpha$-enhanced model (\afe$=+0.4$) and the solar-scaled model.  
The thick black line shows the difference measured from the $\alpha$-enhanced model where both isochrones and stellar spectra are $\alpha$-enhanced. 
The blue line represents the ``spectral effect'', which shows the difference from the model where only stellar spectra are $\alpha$-enhanced. 
In contrast, the red line represents the ``isochrone effects'', showing the difference from the model made with $\alpha$-enhanced isochrones and solar-scaled stellar spectra.

\textbf{$\alpha$-sensitive lines:} 
At fixed \feh, all line indices involving $\alpha$ elements increase with \afe, and the increase is larger for older ages. 
Ca4227 shows a great sensitivity to \afe, mainly driven by the ``spectral effect'', as a result of enhanced Ca abundance. 
In contrast, Ca4455 appears to be less sensitive to the Ca abundance.
This has been pointed out by several previous studies showing that, despite its name, Fe and Cr are the main contributors of this index \citep[e.g.,][]{Tripicco1995, Thomas2003, Lee2009}. 
In \cite{Vazdekis2015}, their $\alpha$-enhanced models at fixed total metallicity show weaker Ca4455 as a result of lower Fe abundance. 
In our models, while enhanced Ca abundance with higher \afe\ slightly increases Ca4455 for older ages (by the spectral effect), its line strength is mostly determined by the isochrone effects.

Magnesium is known to be the most important electron contributor among the $\alpha$ elements \citep[e.g.,][]{Lee2009}, and [Mg/Fe] has been widely used as a proxy of $\rm [\alpha/Fe]$ in local elliptical galaxies to constrain their formation timescales \citep[e.g.,][]{Thomas2003}. 
Indeed, \mgb\ is the most sensitive line to \afe\ in our models. 
While $\alpha$-enhancement in isochrones contributes to the increase in index strength to some extent, the spectral effect dominates. 
\mgb\ is defined as a measure of the combined line strengths of the Mg triplet at $5167, 5172$, and $5183\rm\,\AA$, and one noticeable effect of Mg-enhancement on \mgb\ is the change in line ratios between these lines, particularly the enhancement of $5183\rm\,\AA$ relative to the lines at $5167$ and $5172\rm\,\AA$ (See Fig.~\ref{fig:afsps_individual_lines} in the appendix).  
The $\rm Mg_1$ index does not seem to show a clear trend with \afe. 
One reason could be due to its greater sensitivity to C variations than Mg, as presented in previous studies \citep[e.g.,][]{Korn2005, Coelho2007, Vazdekis2015}. 
For example, \cite{Knowles2023} has also shown that the $\rm Mg_1$ index is insensitive to \afe\ (at constant \feh) for all SSP ages and found that [C/Fe] has a greater impact on this index than \afe\ on a star level. 
The $\rm Mg_2$ index also measures the Mg I triplet with the same bandpass as that of \mgb\ but with different definitions of the pseudo-continuua. 
Thus, this line index shows a similar trend with \afe\ as \mgb. 
While the spectral effect still plays a major role in increasing this index strength, the isochrone effect, which mostly affects the continuum level, becomes more important in this case than in the case of \mgb.

Finally, TiO indices increase with \afe\ as well.  
Both isochrone and spectral effects serve to increase the index strength and contribute almost equally. 
The TiO bands are very sensitive to temperature and are mostly dominated by M-type stars. 
The main effect of $\alpha$-enhancement in isochrones is lowering the temperature. 
In addition, enhancing both Ti and O abundances increases the TiO indices, as can be seen from the spectral effect in Fig.~\ref{fig:afsps_lick_indices_solariso_solarc3k}. 
Interestingly, both $\rm TiO_1$ and $\rm TiO_2$ indices turn over at $\rm \log(age/yr)\sim8.5$ and increase towards younger ages ($<0.1\,\rm Gyr$). 
This behavior has also been found in \cite{Vazdekis2015}, when they used a universal Kroupa IMF, which is attributed to the supergiant phases.
They also found that this behavior appears to be sensitive to the IMF slopes; the high TiO indices at younger ages are not seen when the bottom-heavy IMF is adopted.

\textbf{Fe-sensitive lines:}
At constant \feh, most Fe indices (Fe4383, Fe5270, Fe5335, Fe5406, and Fe5782) decrease with increasing \afe\ for older stellar population ($>1\,\rm Gyr$). 
This is a result of the two opposite effects of $\alpha$-enhancement in isochrones and stellar spectra, as shown in Fig.~\ref{fig:afsps_lick_indices_solariso_solarc3k}. 
$\alpha$-enhancement in isochrones makes the metal lines deeper by lowering the temperature of stars dominating SSP fluxes. 
On the other hand, $\alpha$-enhancement in stellar spectra makes these lines weaker, as found in previous studies \citep[e.g.,][]{Barbuy2003}. 
One possible explanation is that the enhanced abundances of $\alpha$ elements increase the opacity by providing more electrons. As a result, the lines are formed in higher atmospheric layers where the pressure is lower, which could have led to narrower line wings.

However, Fe4531, Fe5015, and Fe5709 indices show increased strengths with \afe\ at fixed \feh. 
The enhanced Fe4531 and Fe5015 indices could be due to their dependence on Ti abundance which is enhanced in our $\alpha$-enhanced models. 
Several studies have shown that the behaviors of these two indices are dominated by the abundance of Ti \citep[e.g.,][]{Tripicco1995, Korn2005, Lee2009, Johansson2012}.
Indeed, Fig.~\ref{fig:afsps_lick_indices_solariso_solarc3k} shows that the spectral effect increases the strengths of these indices. 
The Fe5709 has been known to be less sensitive to Fe abundances than other Fe indices, probably because of its negative sensitivity to carbon \citep[e.g.][]{Tripicco1995}.  
\cite{Korn2005} has found that in high-metallicity giants, this index is more sensitive to Ti abundance than Fe. 
In our models, the increase in this index strength with $\alpha$-enhancement is mainly due to the spectral effect.

Although not one of the iron lines, NaD also shows weaker strengths with increasing \afe\ at constant \feh.
This is consistent with the findings in previous studies showing that NaD behaves as Fe-sensitive lines \citep[e.g.,][]{Coelho2007, CvD12, Vazdekis2015}. 

\textbf{Carbon-sensitive lines:} 
All carbon-sensitive line indices, $\rm CN_2$, G4300, Fe4383 \citep{Tripicco1995}, and $\rm C_2 4668$, show weaker strengths with increasing \afe\ in our models.
This may be due to the enhanced oxygen abundances in the $\alpha$-enhanced models, as enhanced oxygen abundances can cause more carbon to be tied up by the formation of CO, resulting in less carbon for the CN or $\rm C_2$ lines. 
The O-enhancement can also contribute to lowering carbon abundances through the CNO cycle. 
However, as mentioned in Section~\ref{sec:caveats_limitations}, changes in surface abundances as a result of stellar evolution processes (e.g., CNO cycle, dredge-up) are not reflected in our stellar atmosphere calculations; [C/Fe] and [N/Fe] are kept to be solar-scaled ratios throughout all stellar evolutionary phases. Future work is needed to build stellar spectral models that fully reflect the evolution of surface elemental abundances.

\textbf{Age-sensitive lines: }
At \feh$=+0.0$, $\rm H\,\beta$ in the $\alpha$-enhanced models appears to be weaker for younger stellar populations with ages of $\rm \log(age/yr)=8.5-9$ and does not seem to show a clear dependence on \afe\ for older stellar population ($>1\,\rm Gyr$). 
Previous studies have shown that higher-order Balmer series ($\rm H\,\gamma$ and $\rm H\,\delta$) are more sensitive to \afe\ \citep[e.g.,][]{Vazdekis2015, Knowles2023}. 
In Fig.~\ref{fig:afsps_balmer_lines}, we show the age-sensitive line indices ($\rm H\beta$, $\rm H\gamma_A$, $\rm H\delta_A$, and D4000) as a function of SSP ages (zooming in for $\rm \log(age/yr)=8.5-10.3$) for \feh$=-1.0$ (top) and \feh$=+0.0$ (bottom) with different \afe\ (same color-code used as in Fig.~\ref{fig:afsps_lick_indices}). 
The $\rm H\gamma_A$ and $\rm H\delta_A$ are defined in \cite{Worthey1997}, and $\rm D\,4000$ is defined as the flux ratio between [$4050-4250\,\rm \AA$] and [$3750-3950\,\rm \AA$], as in \cite{Bruzual1983, Poggianti1997}. 

We find that $\alpha$-enhanced models at fixed \feh\ show weaker Balmer indices and that the SSP ages sensitive to \afe\ depend on \feh. 
At \feh$=+0.0$, younger stellar populations with ages of $\rm \log(age/yr)=8.5-9$ are the most sensitive to \afe. 
In contrast, at lower metallicity, e.g., \feh$=-1.0$, Balmer lines in older stellar populations (1-10 Gyr) show the greatest sensitivity to \afe. 
Therefore, to accurately derive the ages of stellar populations, a detailed analysis with chemical abundances (\feh,\afe) is required.

In summary, we find that $\alpha$-enhancement at constant \feh\ increases the strength of spectral indices involving $\alpha$ elements (or have a dependence on $\alpha$-abundances) and weakens most other spectral indices. 
This is a combined result of the opposite effects of isochrones and stellar spectra: $\alpha$-enhancement in isochrones lowers the temperature and, thus, increases the strength of the metal lines. 
On the other hand, $\alpha$-enhancement in stellar spectra makes most indices weaker.
However, for the line indices involving $\alpha$ elements, the enhanced $\alpha$ abundances make these lines deeper, resulting in increased strengths for the Ca, Mg, Ti lines (e.g., Ca4227, \mgb, $\rm Mg_2$, and TiO indices), and some Fe lines sensitive to Ti abundances (e.g., Fe4531, Fe5015, and Fe5709).

\subsubsection{The effect of $\alpha$-enhancement at fixed \zh}
\label{sec:results_fixz}

\begin{figure}[hbt!]
    \centering
    \includegraphics[width=1.0\columnwidth]{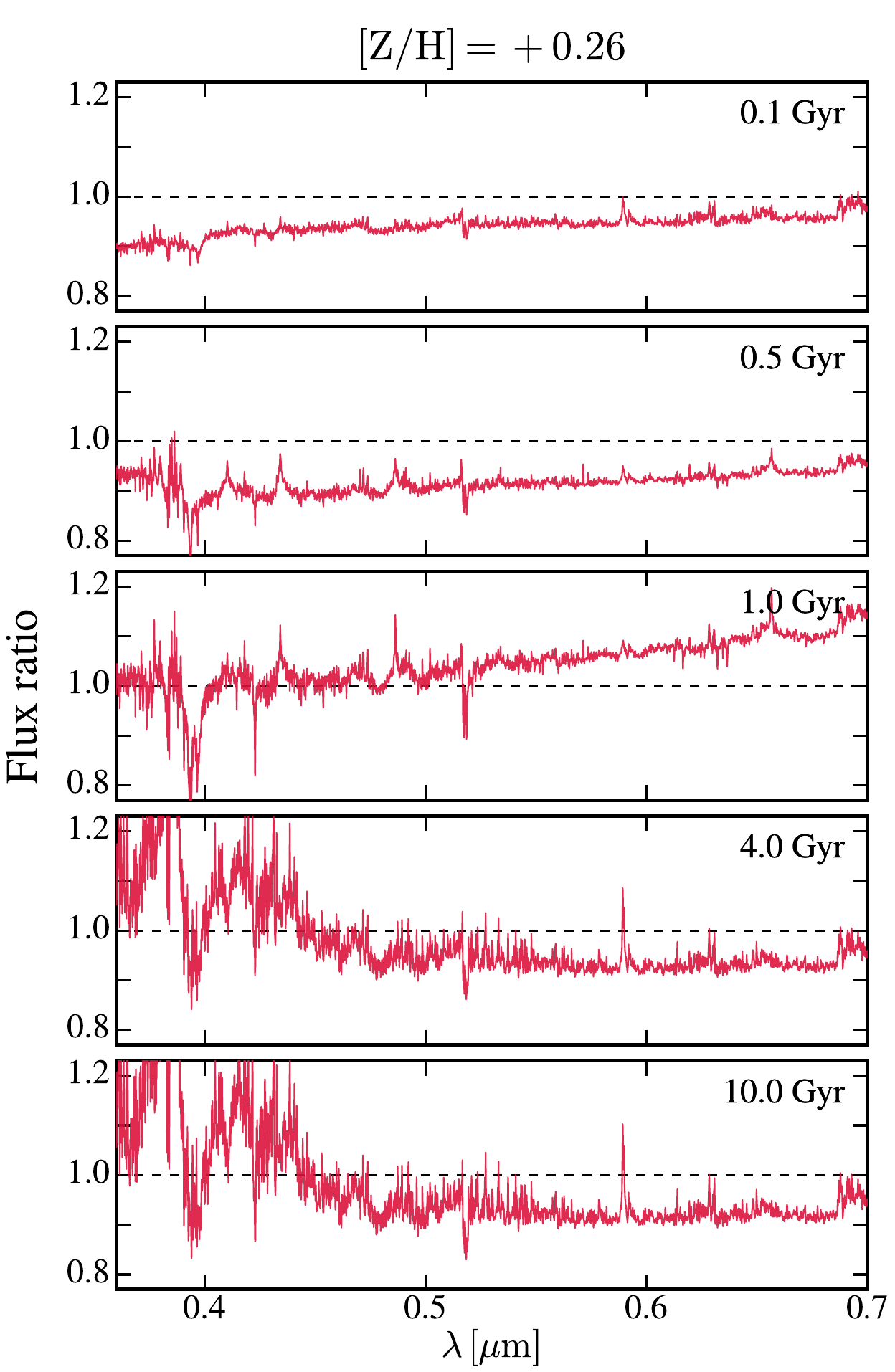}
    \caption{Flux ratios between models with $\alpha$-enhanced (\afe$=+0.4$) and solar-scaled mixtures (\afe$=+0.0$) at constant \zh$=+0.26$. From top to bottom, we show the flux ratios of SSPs for different ages, from 0.1 Gyr to 10 Gyr.}  
    \label{fig:afsps_flux_ratios_fixz}
\end{figure}

\begin{figure}[hbt!]
    \centering
    \includegraphics[width=1.0\columnwidth]{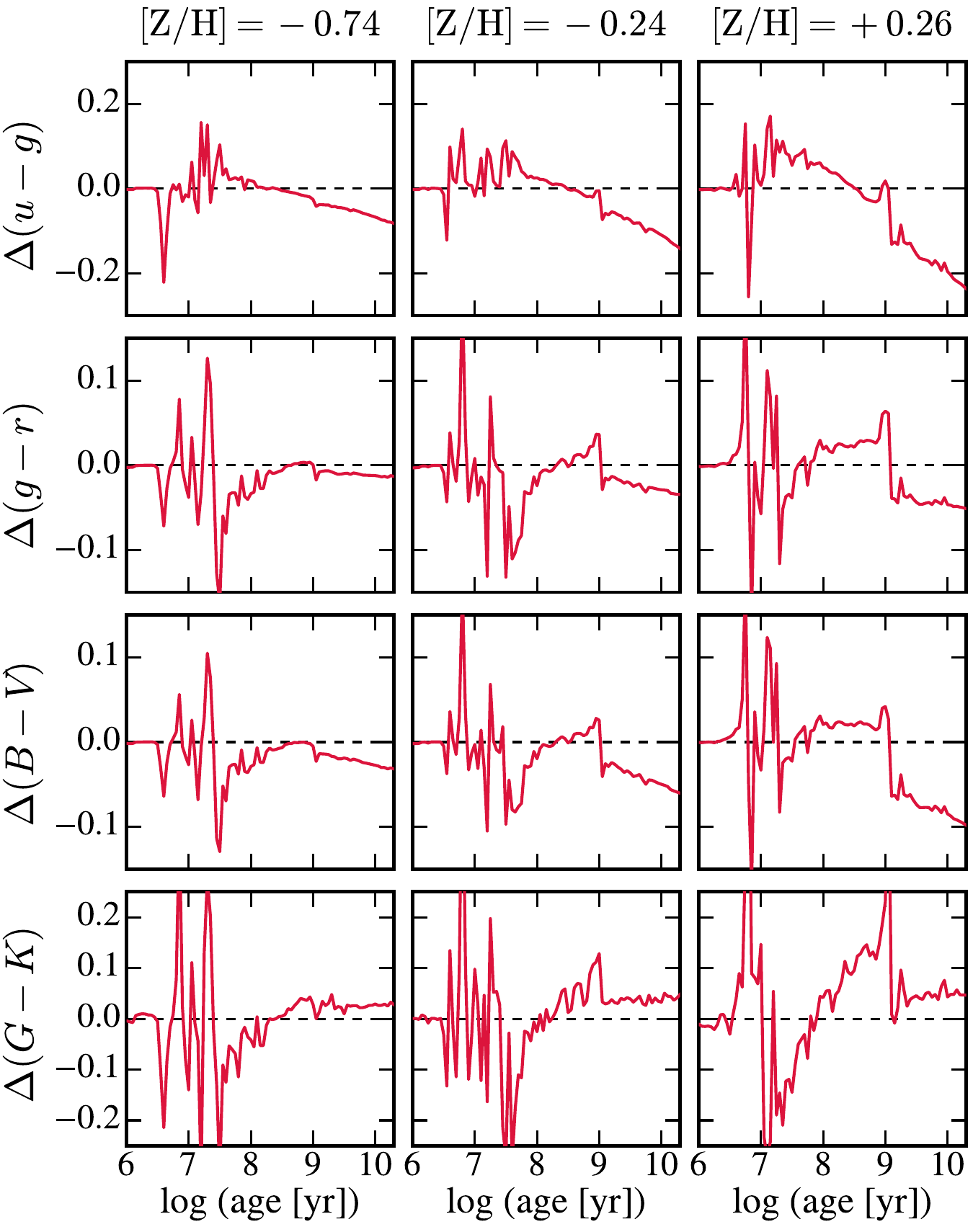}
    \caption{Color differences between the models with $\alpha$-enhanced (\afe$=+0.4$) and solar-scaled mixtures (\afe$=+0.0$) at constant \zh. 
    We find that bluer bands show great sensitivities to the Fe abundances for old stellar populations ($>1\,\rm Gyr$); the models with $\alpha$-enhanced mixtures show much bluer $u-g$ colors due to lower Fe abundances than with the solar-scaled mixtures.}
    \label{fig:afsps_color_fixz}
\end{figure}

\begin{figure*}[hbt!]
    \centering
    \includegraphics[width=1.0\textwidth]{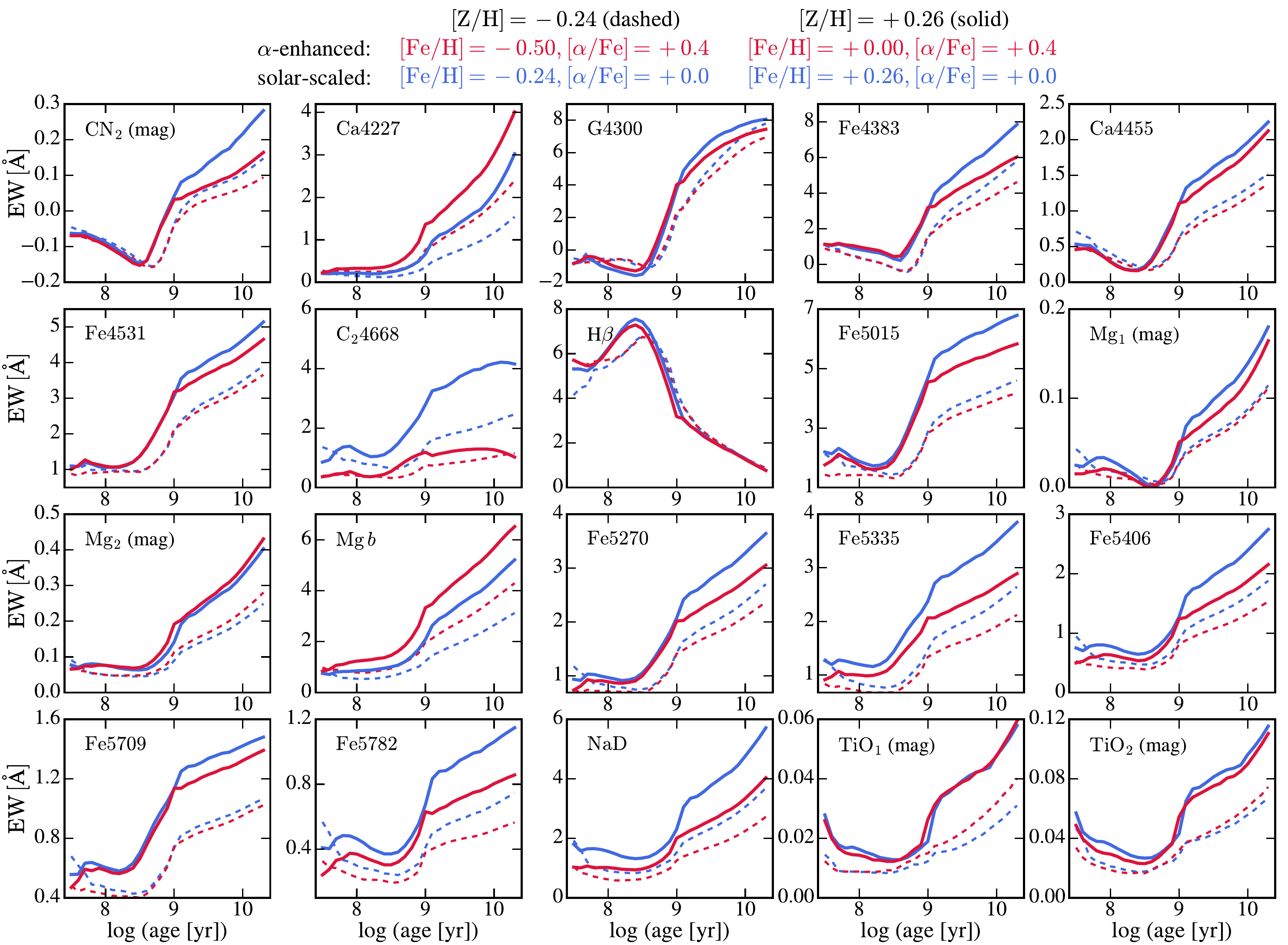}
    \caption{Comparisons of Lick indices between $\alpha$-enhanced (red) and solar-scaled (blue) models at constant \zh: \zh$=-0.24$ (dashed lines) and \zh$=+0.26$ (solid lines). We find that Ca4277 and \mgb\ show great sensitivity to $\alpha$-enhancement. However, all iron (e.g., Fe5015, Fe5270, Fe5335, and Fe5406) and iron-sensitive lines (e.g., Ca4455 and NaD) are weakened in the models with $\alpha$-enhanced mixtures due to decreased iron abundances.}
    \label{fig:afsps_lick_indices_fixz}
\end{figure*}

In this section we explore the effect of $\alpha$-enhancement at constant \zh\ by comparing models having the same \zh\ but with different elemental mixtures: $\alpha$-enhanced mixtures (\afe$=+0.4$) vs. solar-scaled mixtures (\afe$=+0.0$).
At constant \zh, models with $\alpha$-enhanced mixtures have lower Fe abundances than those with solar-scaled mixtures.

Fig.~\ref{fig:afsps_flux_ratios_fixz} shows the flux ratios between models with $\alpha$-enhanced mixtures (\afe$=+0.4$) and solar-scaled mixtures (\afe$=+0.0$) at constant \zh$=+0.26$ for different ages.  
At fixed \zh, models with $\alpha$-enhanced mixtures show lower continuum levels than those with solar-scaled mixtures. 
For older ages ($\rm >1\,Gyr$) and redder wavelengths ($>0.45\,\rm \mu m$), the average flux ratio is $\sim0.9$ and almost constant in the rest-frame optical range.
The flux at bluer wavelengths increases with $\alpha$-enhanced mixtures, leading to color differences in bluer bands.  
For younger stellar populations ($\rm \lesssim 1\,Gyr$), the $\alpha$-enhanced mixtures lower the overall continuum level, especially for shorter wavelengths.

We explore the effect of $\alpha$-enhancement at constant \zh\ on broadband colors. 
Fig.~\ref{fig:afsps_color_fixz} shows color differences between the models with $\alpha$-enhanced (\afe$=+0.4$) and solar-scaled (\afe$=+0.0$) mixtures at constant \zh. 
Each column shows the color differences at each \zh. 
We find that colors at shorter wavelengths show great sensitivities to the Fe abundances for old stellar populations ($>1\,\rm Gyr$); the models with $\alpha$-enhanced mixtures show much bluer $u-g$ colors (by $\sim0.2$ mag for SSPs with $10\,\rm Gyr$) due to lower Fe abundances than those with solar-scaled mixtures.
The bluer $u-g$ colors in the $\alpha$-enhanced mixtures at constant \zh\ are consistent with the results of previous studies \citep[e.g.,][]{Vazdekis2015, Byrne2022}.  
The $g-r$ and $B-V$ colors are also bluer with $\alpha$-enhanced mixtures for old ages ($\rm >1\,Gyr$), but to a much lesser extent. 
The $G-K$ color, on the other hand, is slightly redder (by $\sim0.05$ mag at 10 Gyr) with $\alpha$-enhanced mixtures. 
In Fig.~\ref{fig:afsps_color_evolution}, we found that $\alpha$-enhanced models show much redder $G-K$ color (by $\sim0.2$ mag at 10 Gyr) at fixed \feh. 
This shows that the $G-K$ color, determined by the broad shape of the continuum from the rest-frame optical to near-infrared, is more sensitive to the total metallicity, than to the iron abundances.

Finally, we investigate the effect of $\alpha$-enhancement at constant \zh\ on spectral indices. 
Fig.~\ref{fig:afsps_lick_indices_fixz} compares the Lick indices measured from models with $\alpha$-enhanced (red) and solar-scaled (blue) mixtures at fixed \zh. 
The solid and dashed lines show the models with \zh$=+0.26$ and \zh$=-0.24$, respectively. 
Unsurprisingly, Ca4277 and \mgb\ show great sensitivity to $\alpha$-enhancement. 
All iron and iron-sensitive lines (e.g., Ca4455 and NaD) are weakened in the models with $\alpha$-enhanced mixtures due to decreased iron abundances (Fig.~\ref{fig:afsps_color_fixz}). 
All carbon-sensitive indices also get weakened with $\alpha$-enhanced mixtures at constant \zh.

\subsection{Comparisons to other self-consistent $\alpha$-enhanced models}
\label{sec:result_compare_our_vs_other_afe_models}

\subsubsection{Descriptions of other self-consistent $\alpha$-enhanced models}
\label{sec:other_alpha_models}
In this section we describe other $\alpha$-enhanced SSP models in the literature. Only fully ``self-consistent'' $\alpha$-enhanced models, where both isochrones and stellar spectra are $\alpha$-enhanced, are discussed here. 

\paragraph{\cite{Coelho2007}}
The stellar evolutionary tracks for stars of $0.6-10\,M_\odot$ are computed using the Garching Stellar Evolution code \citep{Weiss2008}. 
The evolutionary sequences are tracked from the MS to the core He-flash for low-mass stars ($M<2\,M_\odot$), up to the early AGB phase for intermediate-mass stars ($M=3-7\,M_\odot$), and up to the beginning of core carbon burning for massive stars ($M=7-10\,M_\odot$).
For the low-mass stars, the post-RGB evolution is extended with the evolutionary tracks from \basti\, models \citep{Pietrinferni2006}. 
The models assume the solar composition from \citetalias{GS98} and uniformly enhance the abundance of all $\alpha$ elements (O, Ne, Mg, Si, S, Ca, and Ti) by 0.4 dex to construct isochrones with \afe$=+0.4$. 
The synthetic spectra with \afe$=+0.0, +0.4$ published by \cite{Coelho2005} are used (covering $3000\rm \AA-1.34\,\mu m$), with an extension to cool giants down to $2800\,\rm K$. 
Both isochrones and synthetic spectra share the same \citetalias{GS98} abundance mixtures. 
The models include SSPs for $3-14\,\rm Gyr$ with six metallicity mixtures: \feh$=-0.5, +0.0, +0.2$ and \afe$=+0.0, +0.4$.

\paragraph{\cite{Lee2009}}
The models have used the Dartmouth $\alpha$-enhanced isochrones \citep{Dotter2007, Dotter2008} where individual elemental variations have been explored. 
Stellar evolutionary sequences are tracked from the MS to the end of RGB. 
At a given total metallicity, 12 sets of stellar evolutionary tracks and isochrones are computed; solar-scaled, $\alpha$-enhanced (all $\alpha$ elements are enhanced simultaneously), and the enhancement of ten individual elements. 
The isochrones are built for six ages: 0.5, 1, 2, 4, 8, and 12 Gyr. 
For stellar spectra, different theoretical stellar spectral models are used for three different temperature regimes. 
The line lists used for the three regimes are not completely homogeneous but share most in common. 
These spectra from three different models are then rebinned into a common wavelength range of $\rm 3000-10000\,\AA$. 
At each ($\log T_{\rm eff}, \log\,g$) grid point, stellar spectra are also computed for 12 different abundance mixtures at fixed $Z$ so that the effect of individual elemental variations on SSPs can be explored self-consistently.

\paragraph{\cite{Vazdekis2015}}
The models used \basti\ $\alpha$-enhanced (\afe=$+0.4$) isochrones \citep{Pietrinferni2006} based on the evolutionary tracks for $0.5-10\,M_\odot$ stars up to carbon ignition or after a few thermal pulses along the TP-AGB phase (depending on the stellar mass). 
The MILES empirical spectral libraries \citep{Sanchez-Blazquez2006} are used to generate base SSP models.  
Differential corrections for $\alpha$-enhancement are computed using the theoretical spectral library of \cite{Coelho2005} and are applied to the base SSPs to synthesize $\alpha$-enhanced SSPs.  
The \mgfe\ estimates of the MILES stars from \cite{Milone2011} are used as a proxy of \afe\ when synthesizing the $\alpha$-enhanced SSPs with differential corrections. 
The $\alpha$-enhanced SSPs have a wavelength range of $\rm 3540-7409\,\AA$ with a resolution of $2.51\,\rm \AA$ (FWHM).

\paragraph{\cite{Knowles2023}; \texttt{sMILES}}
The models used \basti\ $\alpha$-enhanced (\afe=$+0.4$) isochrones and sMILES semi-empiral spectral library \citep{Knowles2021}. 
The sMILES spectral library is based on the MILES empirical stellar spectra with differential corrections made from ATLAS9 model atmospheres \citep{Kurucz1993}. 
The differential correction is applied to individual MILES stars, using \mgfe\ estimates of the MILES stars from \citep{Milone2011} as a proxy of \afe. 
The final sMILES library consists of 801 stellar spectra with \afe$=[-0.2,+0.6]$ in steps of 0.2. 
However, they used solar-scaled isochrones for SSPs with \afe=$-0.2,+0.0,+0.2$ and $\alpha$-enhanced isochrones for SSPs with \afe=$+0.4,+0.6$; therefore, only two \afe\ grids are fully self-consistent: \afe=$+0.0$ and \afe=$+0.4$.
Also, while the solar abundance used for the \basti\ isochrones is from \cite{Grevesse1993}, they use solar abundance from \cite{Asplund2005} to compute the differential correction. 
This inconsistency leads to a complicated relation between the total metallicity [M/H], \feh, and \afe\ of SSPs, which they estimated as follows: $\rm [M/H]_{\rm SSP}=[Fe/H]+0.66[\alpha/Fe]+0.20[\alpha/Fe]^2$ (equation 2 in their paper).  

\vspace{0.5cm}

\begin{figure}[t!]
    \centering
    \includegraphics[width=1.0\columnwidth]{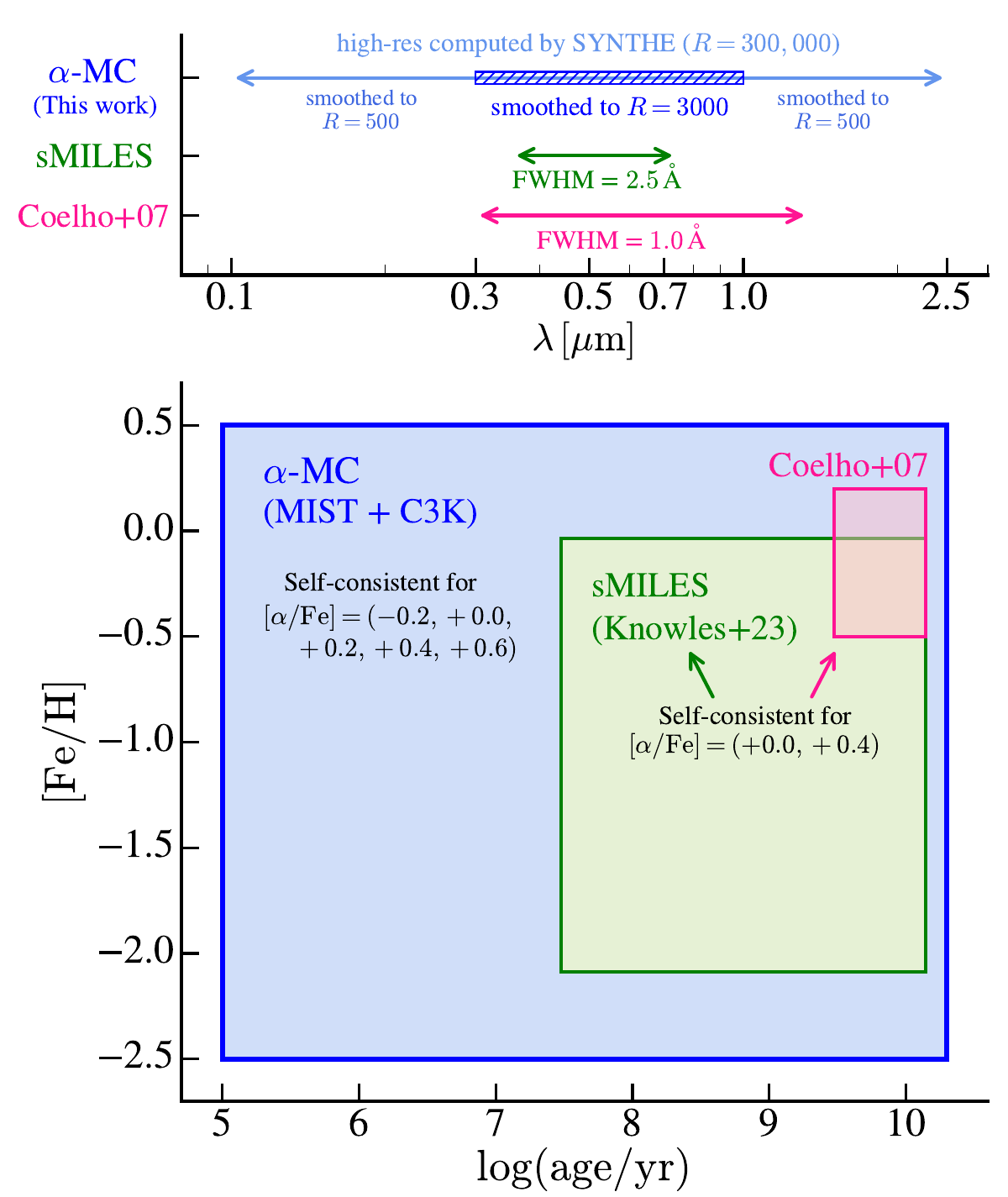}
    \caption{Stellar parameter coverage of self-consistent $\alpha$-enhanced models: wavelength ranges (upper panel) and \feh\ and ages (bottom panel).} 
    \label{fig:model_param_coverage}
\end{figure}

\subsubsection{Comparison results}

\begin{figure*}[hbt!]
\centering
\renewcommand{\arraystretch}{0}
\setlength{\tabcolsep}{0pt}
\begin{tabular}{cc}
\includegraphics[width=1.08\columnwidth]{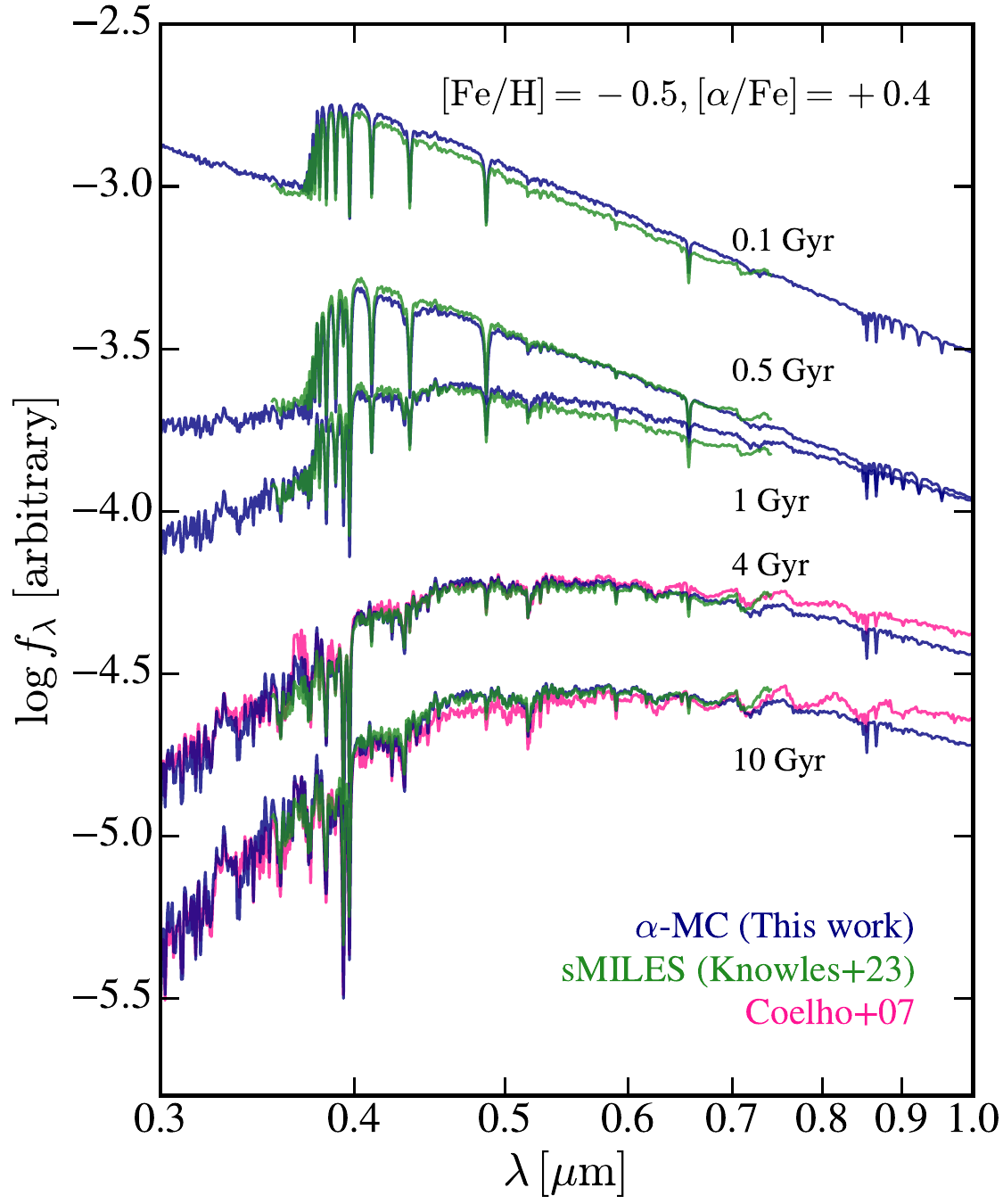} &
\includegraphics[width=1.08\columnwidth]{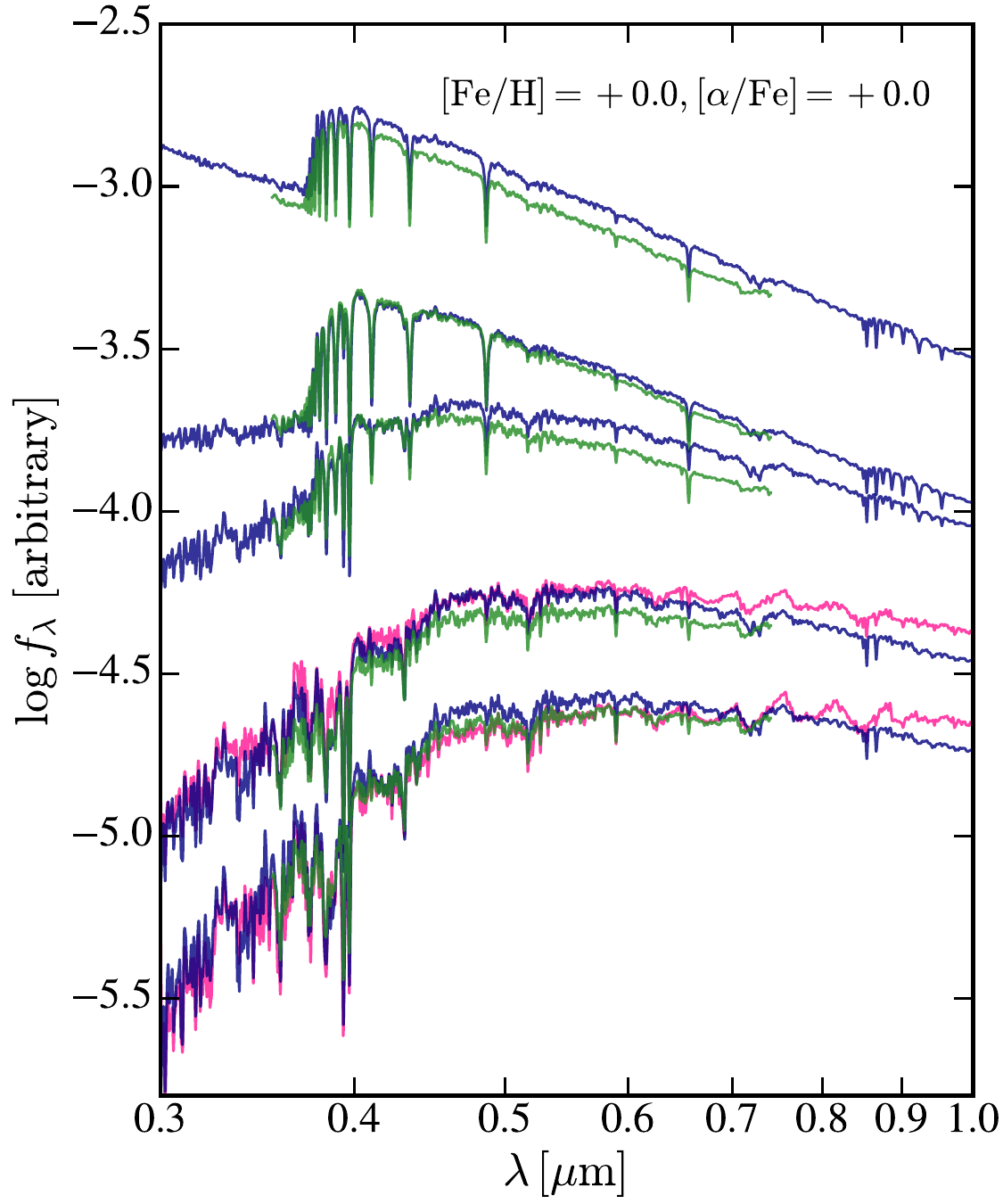}
\end{tabular}
\caption{Model comparison of SSPs for 0.1, 0.5, 1, 4, 10 Gyr with \feh$=-0.5$ and \afe$=+0.4$ (left) and solar-scaled \feh\ and \afe\ (right). }
\label{fig:compare_sed}
\end{figure*}

\begin{figure*}[hbt!]
    \centering
    \includegraphics[width=1.0\textwidth]{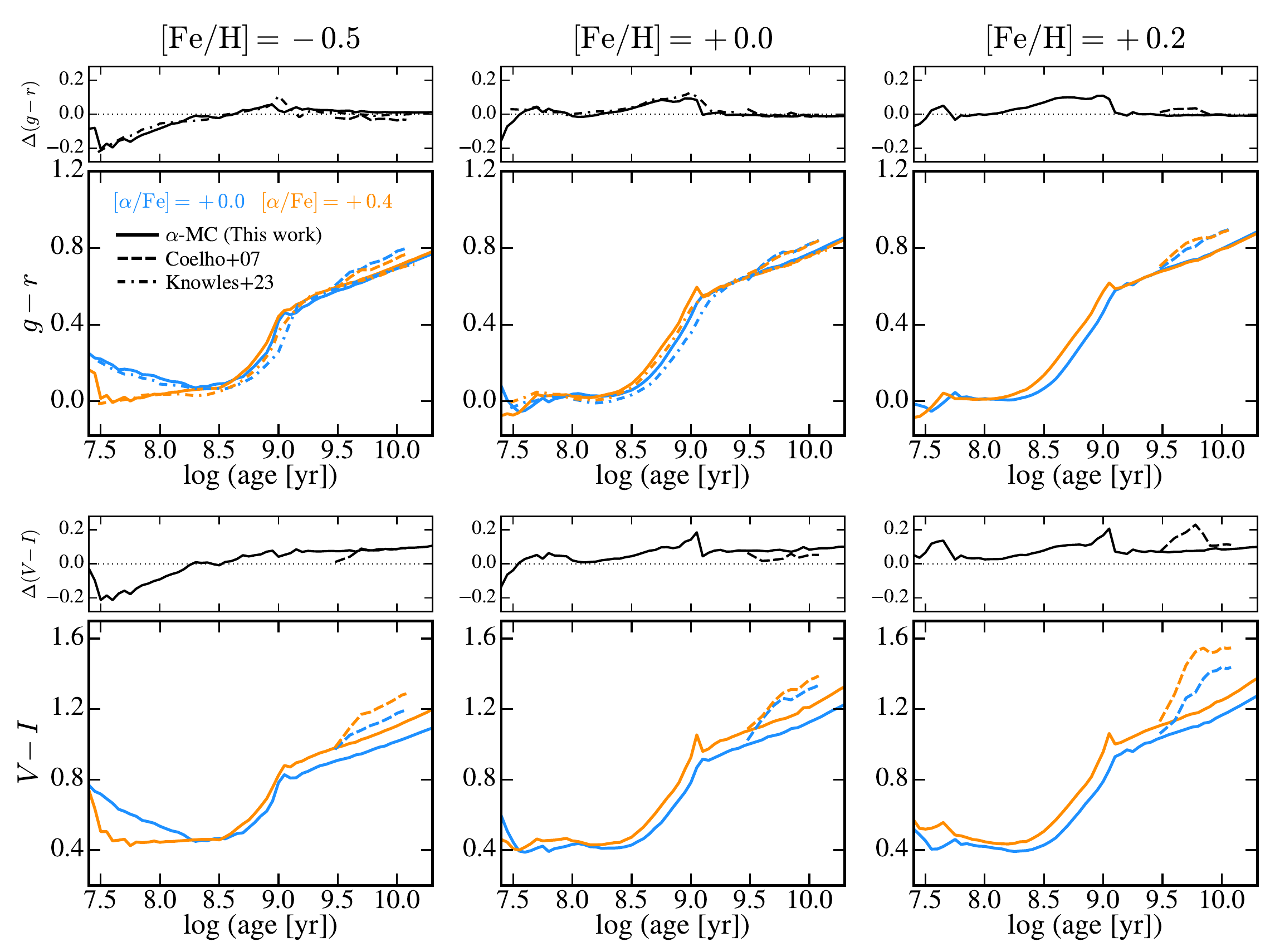}
    \caption{Model comparison of the $g-r$ and $V-I$ color evolution. Blue and orange lines show solar-scaled and $\alpha$-enhanced models (\afe$=+0.4$), respectively. SSPs with \feh$=-0.5, +0.0, +0.2$ are shown in each column. The solid lines show the color evolution of our models. The models from \cite{Coelho2007} and \cite{Knowles2023} are shown as dashed and dot-dashed lines, respectively. In each panel, the top subset panel shows the color difference between the $\alpha$-enhanced (\afe$=+0.4$) and solar-scaled models. The different line types represent different models. }
    \label{fig:compare_color_evolution}
\end{figure*}

\begin{figure*}[hbt!]
    \centering
    \includegraphics[width=1.0\textwidth]{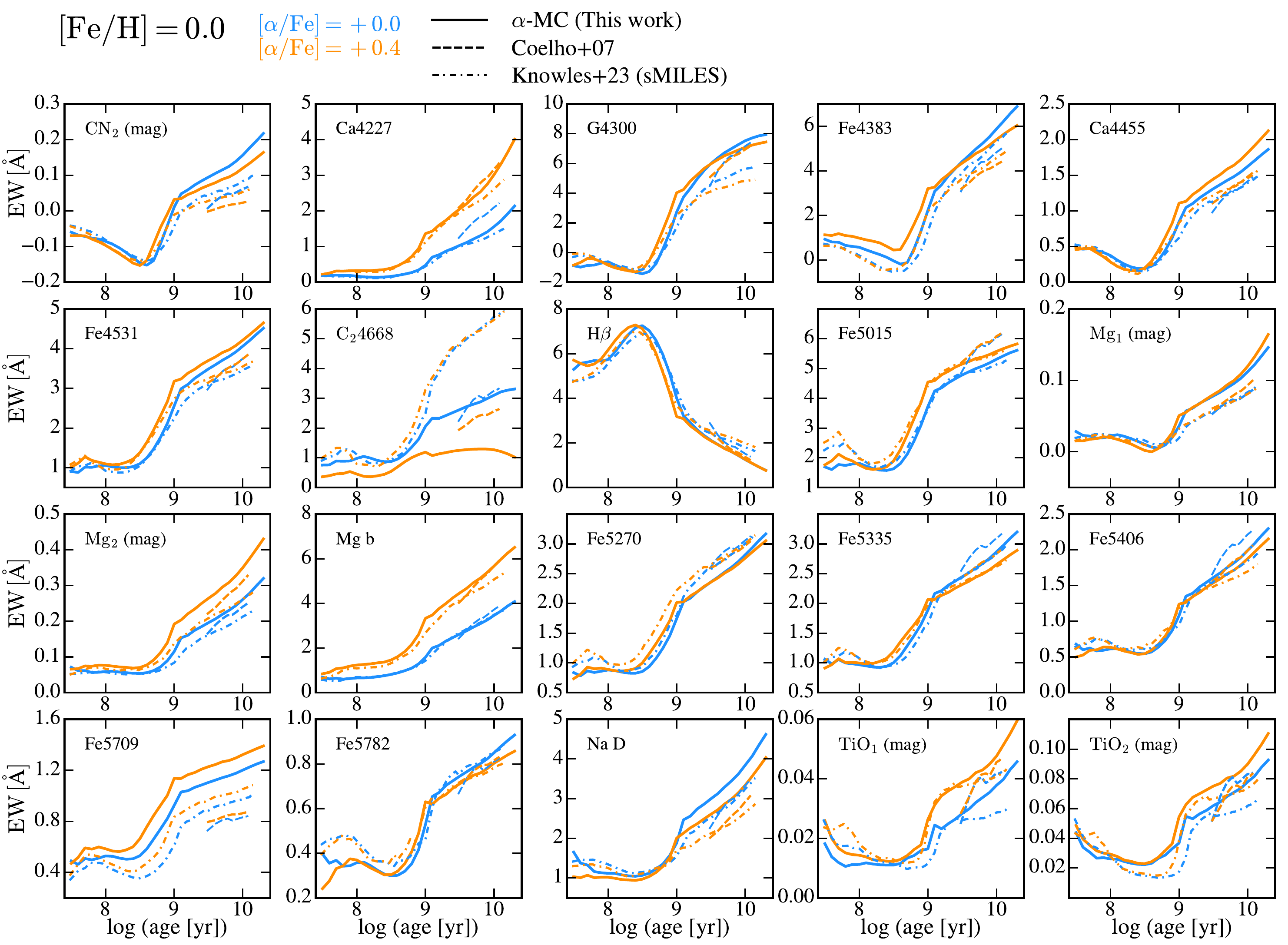}
    \caption{Model comparison of the Lick indices as a function of SSP ages at fixed \feh. The $\alpha$-enhanced (\afe$=+0.4$) and solar-scaled models are shown as orange and blue lines, respectively. 
    The solid lines show the indices measured from our models.
    The models from \cite{Coelho2007} and \cite{Knowles2023} are shown as dashed and dot-dashed lines, respectively. 
    All models are smoothed to $\rm FWHM=8.4\rm \, \AA$ to measure the index strengths, and the Kroupa IMF \citep{Kroupa2001} is assumed for all models. 
    }
    \label{fig:compare_lick_indices}
\end{figure*}

We mainly compare our results with those from \cite{Coelho2007} (using purely theoretical $\alpha$-enhanced spectra) and \cite{Knowles2023} (using semi-empirical $\alpha$-enhanced spectra).  
We do not include the results from \cite{Vazdekis2015}, as the detailed comparisons between \cite{Vazdekis2015} and sMILES SSPs are given in \cite{Knowles2023}. 
Fig.~\ref{fig:model_param_coverage} shows the parameter coverage of the three models in the wavelength range (upper panel) and in the space of \feh\ and age (bottom panel). 
Note that only two \afe\ grids are self-consistent in the \cite{Coelho2007} and \cite{Knowles2023} models: \afe$=+0.0$ and $+0.4$. 
We only compare SSPs with these two \afe\ ratios at fixed \feh\ (\feh$=-0.5$ and $+0.0$) where all three models are available and self-consistent.

Note that \cite{Knowles2023} models have SSP grids with total metallicity [Z/H];
therefore, to compare SSPs at constant \feh, we first find the value of [Z/H] for the corresponding \feh\ and \afe, following the equation (2) in their paper. 
We compute the SSP for a given \feh\ by interpolating the SSPs between the two [Z/H] grids at fixed \afe. 
However, for the grid point of (\feh,\afe)$=(+0.0,+0.4)$, [M/H] is higher than the most metal-rich grid point in their models, so we use their most metal-rich models, [M/H]=+0.26 (corresponding \feh$=-0.037$ and \afe$=+0.4$). 
The \feh\ in Fig.~\ref{fig:model_param_coverage} is derived from their [M/H] grids for \afe$=+0.4$ models.

Fig.~\ref{fig:compare_sed} compares SSPs for 0.1, 0.5, 1, 4, 10 Gyr from different $\alpha$-enhanced models. 
The left panel compares SSPs at \feh$=-0.5$ and \afe$=+0.4$, and the right panel compares solar-scaled SSPs. 
Our models are shown in navy, and the magenta and green lines show the SSP spectra from \cite{Coelho2007} and \cite{Knowles2023}. 
All models are smoothed to $\sigma=250\,\rm km/s$. 
Note that sMILES \citep{Knowles2023} cover only the rest-frame optical ($3540-7409\,\rm \AA$), and \cite{Coelho2007} models only have older ages ($>3\,\rm Gyr$).  
For the older population in the rest-frame optical, where all three models are available, we find an overall agreement in the continuum shapes across the models.

Fig.~\ref{fig:compare_color_evolution} shows the model comparisons of the $g-r$ and $V-I$ color evolution for the solar-scaled (in blue lines) and $\alpha$-enhanced (in orange lines) SSPs at \feh$=-0.5, +0.0, +0.2$. 
The solid lines show the color evolution from our models.
The dashed and dot-dashed lines represent the color evolution from \cite{Coelho2007} and \cite{Knowles2023} models, respectively. 
Note that the sMILES semi-empirical spectral library used in \cite{Knowles2023} has a wavelength range of $3540.5 - 7409.6\,\rm \AA$.
For the $V-I$ color, we only compare to \cite{Coelho2007} models as $I$-band is outside the wavelength range of the \cite{Knowles2023} models.
Also, for the \feh$=+0.0$ \afe=$+0.4$ model, [M/H] is higher than the most metal-rich grid in \cite{Knowles2023}, so we use the [M/H]=+0.26 grid (corresponding to \feh$=-0.037$ and \afe$=+0.4$) and do not include sMILES models for \feh$=+0.2$. 
Note that \cite{Coelho2007} models cover only $3000\rm \AA-1.34\,\mu m$, thus, $I$-band is the reddest band we can compare with our models.

Broadly, we find a good overall agreement in the $g-r$ color evolution in these three models. 
The $\alpha$-enhanced and solar-scaled models show the largest differences at $\rm \log(age/yr)\sim9$ for both our models and \cite{Knowles2023} ($\sim0.1$ mag). 
For $V-I$ color, the $\alpha$-enhanced models (at constant \feh) show redder $V-I$ color (by $\sim0.1$ mag) in both our models and \cite{Coelho2007} models. 
The $V-I$ color in the \cite{Coelho2007} models are redder than our models for both $\alpha$-enhanced and solar-scaled models. 
This might be related to inconsistent stellar evolutionary models used in \cite{Coelho2007} where post-RGB evolution (beyond He flash) are stitched with the evolutionary tracks from \texttt{BaSTI}.

Fig.~\ref{fig:compare_lick_indices} shows the model comparisons of the Lick indices as a function of ages. 
All SSP models have constant \feh$=+0.0$, and the solar-scaled and $\alpha$-enhanced (\afe=$+0.4$) models are shown as blue and orange lines, respectively. 
The indices measured from our models are shown in solid lines. 
The dashed and dot-dashed lines show the indices measured from \cite{Coelho2007} and \cite{Knowles2023} models, respectively. 
All models are smoothed to $\rm FWHM=8.4\rm \AA$ to measure the indices and assume the Kroupa IMF \citep{Kroupa2001}.

Overall, the trend with $\alpha$-enhanced models compared to solar-scaled models seems consistent in these models; at constant \feh, indices involving $\alpha$-elements (or have dependence on $\alpha$-elements) increase with \afe, while the other indices get weakened with increasing \afe.

The $\rm C_2\,4668$ index appears to show a large discrepancy between the models using theoretical spectra (our models and \citealt{Coelho2007}) and empirical spectra \citep{Knowles2023}. 
For solar-scaled models, $\rm C_2\,4668$ measured with our models looks consistent with that of \cite{Coelho2007}. 
However, in our models, this index dramatically decreases with increasing \afe\ for older stellar populations ($>1\,\rm Gyr$). 
The large discrepancy in this index between theoretical and empirical spectra has been discussed in \cite{Knowles2019}. 
They have compared the Lick indices of cool dwarf and giant stars measured from synthetic spectra of \cite{Conroy2012}, \cite{Coelho2014}, and \cite{AllendePrieto2014} models with those from MILES empirical spectra.
They found that all these theoretical spectral models show weaker strength of $\rm C_2\,4668$ compared to the MILES spectra.
As mentioned in Section~\ref{sec:caveats_limitations}, this could be related to that the evolution of surface C and N abundances throughout the stellar evolutionary phases are not reflected in the atmosphere and radiative transfer calculations in our models (so as in \cite{Coelho2007}). 
Another possibility could be the theoretical line list used in our models; an updated line list might help improve the prediction for this index strength. 

\vspace{3cm}
\section{Summary and conclusion}
\label{sec:summary_conclusion}

In this work we presented \amc, new self-consistent stellar population models with non-solar \afe\ abundance constructed from $\alpha$-enhanced \mist\ isochrones and \ctk\ theoretical spectral library. 
Our new SSP models cover a wide range of ages ($\rm \log (age/yr) = 5.0 - 10.3$), metallicities (\feh$=[-2.5,+0.5]$ in steps of 0.25, \afe$=-0.2,+0.0,+0.2,+0.4,+0.6$), and wavelengths (a high-resolution is computed for $\rm 0.1-2.5\,\mu m$, smoothed to a resolution of $R=3000$ for $\rm 0.3-1.0\,\mu m$ and to a lower resolution outside this range). 
The $\alpha$-enhanced isochrones and stellar spectra share the same abundance mixtures from \citetalias{GS98}, therefore, our SSPs are fully consistent in terms of the abundances. 
We explored the effect of $\alpha$-enhancement on SSP full spectra, broadband colors, and spectral indices. 
We further separated the ``isochrone effect'' and ``spectral effect'' of $\alpha$-enhancement and investigated each effect by generating models in which only isochrones or stellar spectra are $\alpha$-enhanced. 
We also compare our models with other previous self-consistent $\alpha$-enhanced models, in terms of SSP broadband colors and spectral indices, for the overlapping coverage. 
Here we summarize our results:

\begin{enumerate}
    \item We find that the primary effect of $\alpha$-enhancement of isochrones on SSP spectra is to lower the overall continuum and redden the continuum shapes, which is more pronounced for higher \feh\ and older SSP ages. 
    For \feh$=+0.0$, SSPs with $\rm [\alpha/Fe]=+0.4$ have lower overall continuum levels by $\sim20\%$. 
    On the other hand, $\alpha$-enhancement in stellar spectra mainly affects the individual spectral lines.

    \item $\alpha$-enhancement affects broadband SSP colors across all ages. 
    At fixed \feh, SSPs with \afe$=+0.4$ have redder $G-K$ color by $\sim0.2$ mag for ages $>1\,\rm Gyr$. 
    For old ages ($>1\,\rm Gyr$), the effect of $\alpha$-enhancement (at fixed \feh) on colors at shorter wavelengths is more subtle, due to the opposite effects of isochrones (making redder by lower temperature of stars dominating the light) and stellar spectra (making bluer by increasing fluxes at blue wavelengths). 
    Instead, colors at shorter wavelengths appear to be more sensitive to iron abundances than to the total metallicity. 
    At constant \zh, the models with $\alpha$-enhanced mixtures (with lower Fe abundances) show much bluer $u-g$ color (by $\sim0.2$ mag) for old stellar populations ($>1\,\rm Gyr$). 
    $\alpha$-enhancement shows huge impacts on colors for young stellar populations ($\rm 0.01-1\,Gyr$), associated with stellar evolutionary phases (e.g., the appearance of red super giants or TP-AGB phases).

    \item $\alpha$-enhancement has significant impacts on SSP spectral indices. 
    At constant \feh, $\alpha$ enhancement increases the strength of spectral indices involving $\alpha$-elements (e.g., Ca4227, \mgb, and TiO indices) and weakens most other iron- and carbon-sensitive lines (e.g., Fe5270, Fe5335, and NaD). 
    This is a combined result of the opposite effects of isochrones and stellar spectra: $\alpha$-enhancement in isochrones increases the strengths of metal lines by lowering the temperature of stars dominating the fluxes. 
    On the other hand, $\alpha$-enhancement in stellar spectra makes most indices weaker.  
    At constant \zh, models with $\alpha$-enhanced mixtures show stronger Ca4227 and \mgb.  
    However, all iron (e.g., Fe5015, Fe5270, Fe5335, and Fe5406) and iron-sensitive indices (e.g., Ca4455 and NaD) are weakened with $\alpha$-enhanced mixtures due to lower Fe abundances.

    \item At constant \feh, $\alpha$-enhanced models show weaker Balmer strengths than solar-scaled models.
    The SSP ages sensitive to \afe\ also depends on \feh. 
    This shows that a detailed analysis with chemical abundances is needed to accurately derive the age of stellar populations.

\end{enumerate}

In conclusion, we find that \textit{$\alpha$-enhancement in isochrones and stellar spectra have opposite effects on SSP colors and spectral indices}. 
Because the effect of $\alpha$-enhancement on SSPs is a combination of these two opposite effects, \textit{$\alpha$-enhancement should be taken into account, self-consistently, for both isochrones and stellar spectra.} 
In this paper, we presented our new \amc\ models, explored the effects of $\alpha$-enhancement on SSPs, and compared them with other self-consistent $\alpha$-enhanced models. 
In the subsequent paper, we will present our results applying the \amc\ models to observed data.

The advent of JWST has pushed the frontier of the high-redshift Universe and revealed a surprisingly abundant population of early galaxies. 
In the early universe, SNe II were more dominant for chemical enrichment, thus, galaxies are expected to be $\alpha$-enhanced. 
To derive accurate physical properties of these high-$z$ galaxies, fully self-consistent $\alpha$-enhanced stellar population models are required.
Our new $\alpha$-MC models, including young stellar populations ($<1\,\rm Gyr$) and wide wavelength coverage (rest-frame UV, optical, and near-infrared), will be essential in deriving accurate physical properties of high-redshift galaxies, where $\alpha$-enhancement is expected to be common.

\begin{acknowledgments}
This project was funded in part through NSF grant AST-1908748. 
The computations in this paper were run on the FASRC Cannon cluster supported by the FAS Division of Science Research Computing Group at Harvard University. 
\end{acknowledgments}

\clearpage
\appendix
\renewcommand{\thefigure}{A\arabic{figure}}
\setcounter{figure}{0}

\renewcommand{\thetable}{A\arabic{table}}
\setcounter{table}{0}

\section{Spectral indices}

\begin{figure*}[hbt!]
    \centering
    \includegraphics[width=1.0\textwidth]{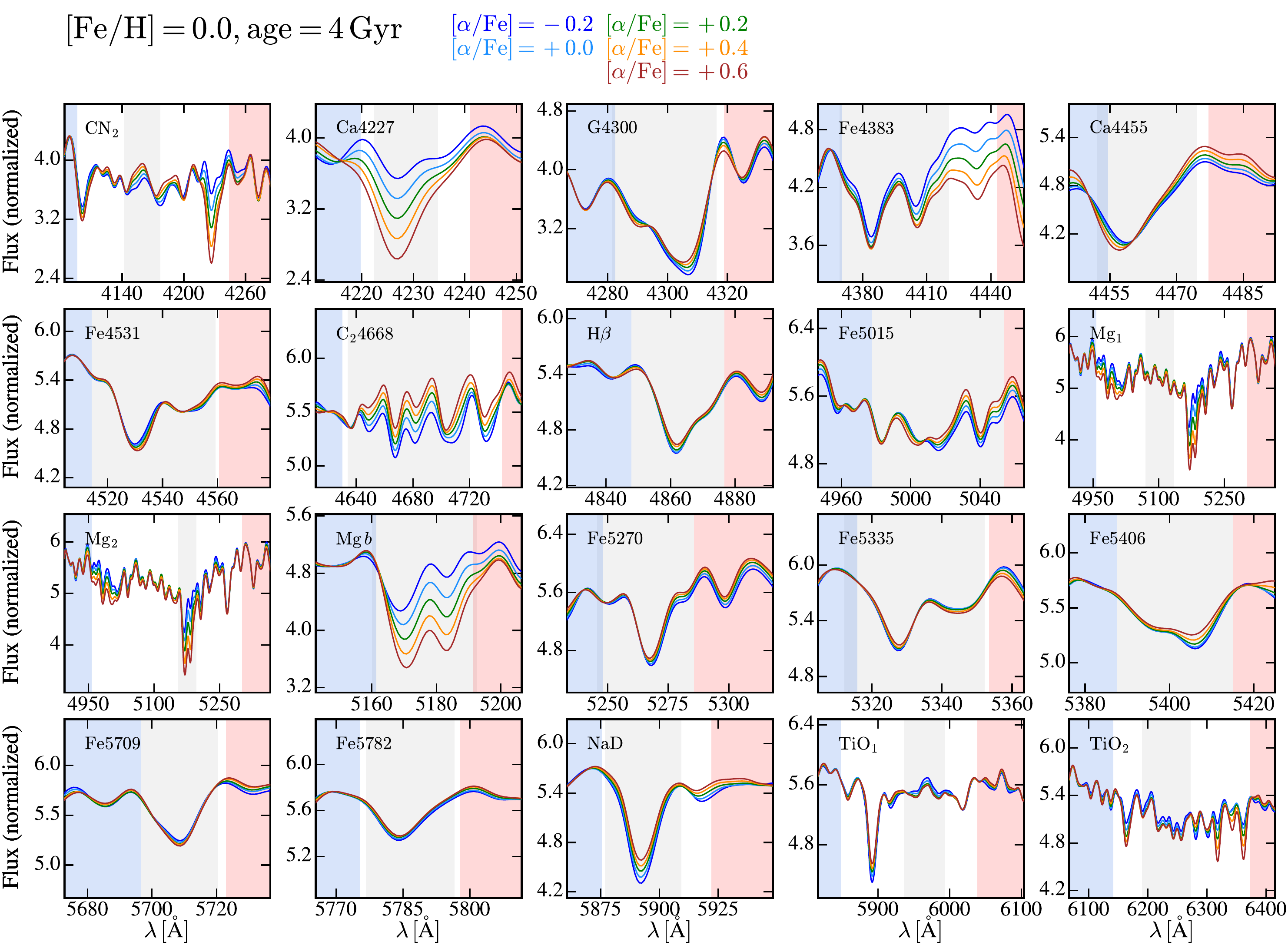}
    \caption{SSPs around individual Lick indices with \feh$=+0.0$ and varying \afe\ (shown as different colors) for fixed age of 4 Gyr. Each panel zooms in around each Lick index. The gray, blue, and red shades indicate the index bandpass, blue and red pseudo-continua defined in \cite{Trager2000}. The SSP spectra are smoothed to $8.4\,\rm \AA$ (FWMH) resolution. 
    The flux of SSP spectra with varying \afe\ is normalized to the midpoint of the blue continuum of solar-scaled SSP spectra.}
    \label{fig:afsps_individual_lines}
\end{figure*}

Fig.~\ref{fig:afsps_individual_lines} shows SSPs around individual Lick indices with \feh$=+0.0$ and varying \afe\ (shown as different colors) for a fixed age of 4 Gyr. Each panel zooms in around each Lick index. 
The gray, blue, and red shades indicate the index bandpass, blue and red pseudo-continua defined in \cite{Trager2000}. The SSP spectra are smoothed to $8.4\,\rm \AA$ (FWMH) resolution. 
The fluxes of SSP spectra with varying \afe\ are normalized to the midpoint of the blue continuum of solar-scaled SSP spectra. 
We measure the Lick indices from SSP spectra at a resolution of $8.4\,\rm \AA$ (FWHM).

\clearpage

\bibliography{references}{}
\bibliographystyle{aasjournal}
\end{document}